\documentclass[fleqn,usenatbib]{mnras}

\usepackage{newtxtext,newtxmath}
\usepackage{url}
\usepackage{color}
\usepackage{tabulary}
\usepackage{multirow}
\usepackage{amsfonts}
\usepackage{array}
\usepackage{ulem}
\usepackage{pdflscape}
\usepackage[T1]{fontenc}
\usepackage{ae,aecompl}

\usepackage{graphicx}	
\usepackage{amsmath}	
\usepackage{amssymb}	
\usepackage{cases}

\graphicspath{ {pictures/} }

\newcommand{\beq}{\begin{equation}}
\newcommand{\eeq}{\end{equation}}
\newcommand{\simlt}{\mathrel{\hbox{\rlap{\hbox{\lower4pt\hbox{$\sim$}}}\hbox{$<$}}}}
\newcommand{\simgt}{\mathrel{\hbox{\rlap{\hbox{\lower4pt\hbox{$\sim$}}}\hbox{$>$}}}}

\newcommand{\erg}{\;\mathrm{erg}}
\newcommand{\s}{\;\mathrm{s}}

\newcommand{\gram}{\;\mathrm{g}}
\newcommand{\cm}{\;\mathrm{cm}}

\newcommand{\cmcube}{\;\mathrm{cm}^{-3}}
\newcommand{\km}{\;\mathrm{km}}

\newcommand{\K}{\;\mathrm{K}}
\newcommand{\Pbar}{\;\mathrm{bar}}
\newcommand{\Pmbar}{\;\mathrm{mbar}}

\def\apjl{ApJL}
\def\apj{ApJ}
\def\mnras{M.N.R.A.S.}
\def\aap{A\&A}

\def\apjs{ApJ Supp.}

\title[Hot Jupiter]{Turbulence-driven thermal and kinetic energy fluxes in the atmospheres of hot Jupiters}

\author[T. Ryu et al.]{
Taeho Ryu$^{1}$\thanks{email: taeho.ryu@stonybrook.edu}, 
Michael Zingale$^{1}$,
Rosalba Perna$^{1}$\\ 
$^{1}$Department of Physics and Astronomy, Stony Brook University, Stony Brook, NY 11794-3800, USA}

\date{Accepted XXX. Received YYY; in original form ZZZ}

\pubyear{2018}

\begin{document}
\label{firstpage}
\pagerange{\pageref{firstpage}--\pageref{lastpage}}
\maketitle

\begin{abstract}
	We have performed high resolution $3-$dimensional compressible hydrodynamics simulations to investigate the effects of shocks and turbulence on energy transport into hot Jupiter atmospheres, under a variety of shear gradients. We focus on a local atmospheric region to accurately follow the small-scale structures of turbulence and shocks. We find that the effects of turbulence above and below a shear layer are different in scale and magnitude: below the shear layer, the effects of turbulence on the vertical energy transfer are local, generally $\lesssim2~\times(\text{scale height})$. However, turbulence can have a spatially and thermally-large influence on almost the entire region above the shear layer. We also find that shock formation is local and transient.
Once the atmosphere becomes steady, the time-averaged heat  flux at $P\sim 1\Pbar$ is insignificant, on the order of 0.001\% of the incoming stellar flux with a shear motion at $P\simeq 1\Pmbar$, and 0.1\% with a deeper shear layer at $P\simeq 100\Pmbar$. Accordingly, the diffusion coefficient is higher for the deeper shear layer. Therefore, our results suggest that turbulence near less dense ($P\simeq 1\Pmbar$) regions does not cause a sufficiently deep and large penetration of thermal energy to account for radius inflation in hot Jupiters, regardless of how violent the turbulence is. However, as the shear layer gets deeper, heat transfer becomes more effective throughout the atmosphere (upwards and downwards) due to a larger kinetic energy budget. Therefore, it is more important how deep turbulence occurs in the atmosphere, than how unstable the atmosphere is for effective energy transfer. 

\end{abstract}

\begin{keywords}
Hot Jupiter $-$ planetary systems : atmosphere $-$ planetary systems : gaseous planet
\end{keywords}



\section{Introduction}
\label{sec:intro}

Hot Jupiters are a class of gas-giant exoplanets, characterized by short orbital periods ($P\lesssim50$ days). Such close proximity to their parent stars leads to several interesting features, which include tidal synchronization, strong irradiation, and a generally large day-night temperature contrast. A number of hot Jupiters are observed to have radii larger than what predicted from standard cooling models \citep[e.g.][]{Showman2002,Guillot2002,Howard+2012,Wang+2015}. The origin of the radius inflation is still debated, and several ideas have been put forward to explain it.

Inflated radii imply that the bloated planets retain more internal entropy than expected. This could be produced by either injection/dissipation of heat, or less efficient energy loss, or a combination of both. Within this context, the mechanisms that have been put forward to explain the radius anomaly can be divided into two classes. The first category includes less efficient cooling due to enhanced opacity \citep{Burrows+2007}. As the opacity increases, cooling becomes inefficient and the planets can naturally retain more internal heat. The second category invokes extra heat sources in the interior, such as the dissipation of heat via tidal forces  \citep{Bodenheimer+2001,Jackson+2008,IbguiBurrows2009,Ibgui+2011}, conversion of the stellar flux into kinetic energy of the global atmospheric flow, driven by the large day-night temperature gradient (often called ``hydrodynamic dissipation"; \citealt{Showman2002,Guillot2002,Showman+2009,Heng+2011b,Heng+2011a}), magnetic drag in ionized planetary winds, or ``ohmic dissipation" \citep{BatyginStevenson2010,Perna+2010a,Perna+2010b,Perna+2012}, and dissipation of energy induced by fluid instabilities \citep{LiGoodman2010}. For a comprehensive comparison and review, see \citet{HengShowman2015}.

Among those, energy dissipation via turbulence \citep{LiGoodman2010}, likely accompanied by shocks \citep{Perna+2012,Dobbs-DixonAgol2013,Heng2012}\footnote{Generally, fluid in a stably stratified atmosphere becomes unstable when the shear stress, or velocity gradient, is sufficiently large that buoyancy forces suppress the vertical displacements of fluid elements. Therefore, high-speed flows are more likely subject to the instability.}, could be a viable, or at least interesting mechanism to consider. This is because turbulence may be ubiquitous and present even in stably stratified atmospheres. It is hence natural to study its onset in globally circulating planetary atmospheres. 
\citet{YoudinMitchell2010} proposed that forced turbulence can drive downward transport of heat in the outer radiative zone of stratified atmospheres. They called this the ``mechanical greenhouse effect" \footnote{See \citet{Izakov2001} for the greenhouse effect in the atmosphere of Venus.}, and built an analytic model of the outer radiative zone, focusing on diffusion and dissipation by forced turbulence. They found that a heat flux generated by forced turbulence propagates downwards and can be deposited in deeper regions. Their analytic approach, undoubtedly necessary for understanding the underlying physics, is however more suitable for somewhat idealized cases. To account for a more realistic scenario, simulations with detailed modelling are essential.  Recently, using the compressible shock-capturing code RAMSES, \citet{Fromang+2016} developed a 3-dimensional model to examine the role of shear-driven instabilities and shocks in planetary atmospheres, using a  Newtonian relaxation scheme. They covered a large volume of the atmosphere to take into account global motions and included cooling via a Newtonian cooling method. Their simulations suggest that equatorial jets are subject to shear-driven instabilities, which can lead to a sufficiently large amount of downward kinetic energy flux and the formation of shocks at a few mbar pressure levels. Their results improve and deepen our understanding of the physics of turbulence and shocks. However, as they pointed out in their paper, it is possible that their spatial resolution may still be too large to capture processes occurring on small scales. 

Motivated by those studies and in order to improve on some of their limitations, in this work we investigate the effect of shocks and turbulence on energy penetration in stable stratified atmospheres, using high resolution 3-dimensional compressible hydrodynamics simulations with the adaptive-mesh finite-volume code CASTRO \citep{Almgren+2010}. We focus on a local atmospheric region to accurately capture the small-scale structures of the eddy motion. We estimate how much and how deep heat can be deposited in the atmosphere when shear motions are driven. Based on the measured heat flux, we further estimate the diffusion coefficient $K_{\rm zz}$ (see Equation \ref{eq:heatflux_Kzz}). Last, we discuss the formation, duration and distribution of shocks in the planetary atmospheres.

In our suites of simulations, we find that the effects of turbulence on the kinetic and heat energy transfer are local, generally confined to within a spatial range of $z\sim2H$ (where $H$ is the scale height) below where eddies are created, but turbulence can make a spatially and thermally large-scale impact on the regions above it. We also find that shock formation is local and transient.
The time-averaged heat energy flux at $P\sim 1\Pbar$ when the atmosphere becomes steady is on the order of $0.1-001\%$ of the incoming stellar flux depending on the location of the shear layer (lower flux for an outer shear layer). Hence, our results suggest that turbulence near less dense regions ($P\gtrsim 1\Pmbar$) does not lead to a sufficient amount of thermal energy burial in deeper regions to account for the inflated radii of hot Jupiters, regardless of how violent the turbulence is. On the other hand, thermal energy can be transferred more effectively throughout the atmosphere when turbulence is triggered at deeper regions ($P\gtrsim 100\Pmbar$). Therefore, it is more important how deep turbulence occurs in the atmosphere, than how unstable the atmosphere is for effective transfer of energy.

This paper is organized as follows. In Section \ref{sec:numericalsetup}, we explain our numerical setup including the model description (Section \ref{sec:modeldescription}) and the boundary conditions (Section \ref{sec:boundarycondition}), and describe our shear prescription (Section \ref{sec:shearprescription}) and initial model parameters (Section \ref{sec:initialcondition}). We present our results in Section \ref{sec:results}. In Section \ref{sec:discussion}, we first compare our results with two different numerical resolutions for the same set-up, and then we compare simulations at higher resolution but with different atmospheric depths for the shear layer. Finally, we conclude with a summary of our findings in Section \ref{sec:summary}.

\section{Numerical Setup}\label{sec:numericalsetup}

In this section we present our planetary atmosphere models. We describe the initial conditions of the model atmospheres and our shear prescription.

\begin{figure*}
	\centering
	\includegraphics[width=8.4cm]{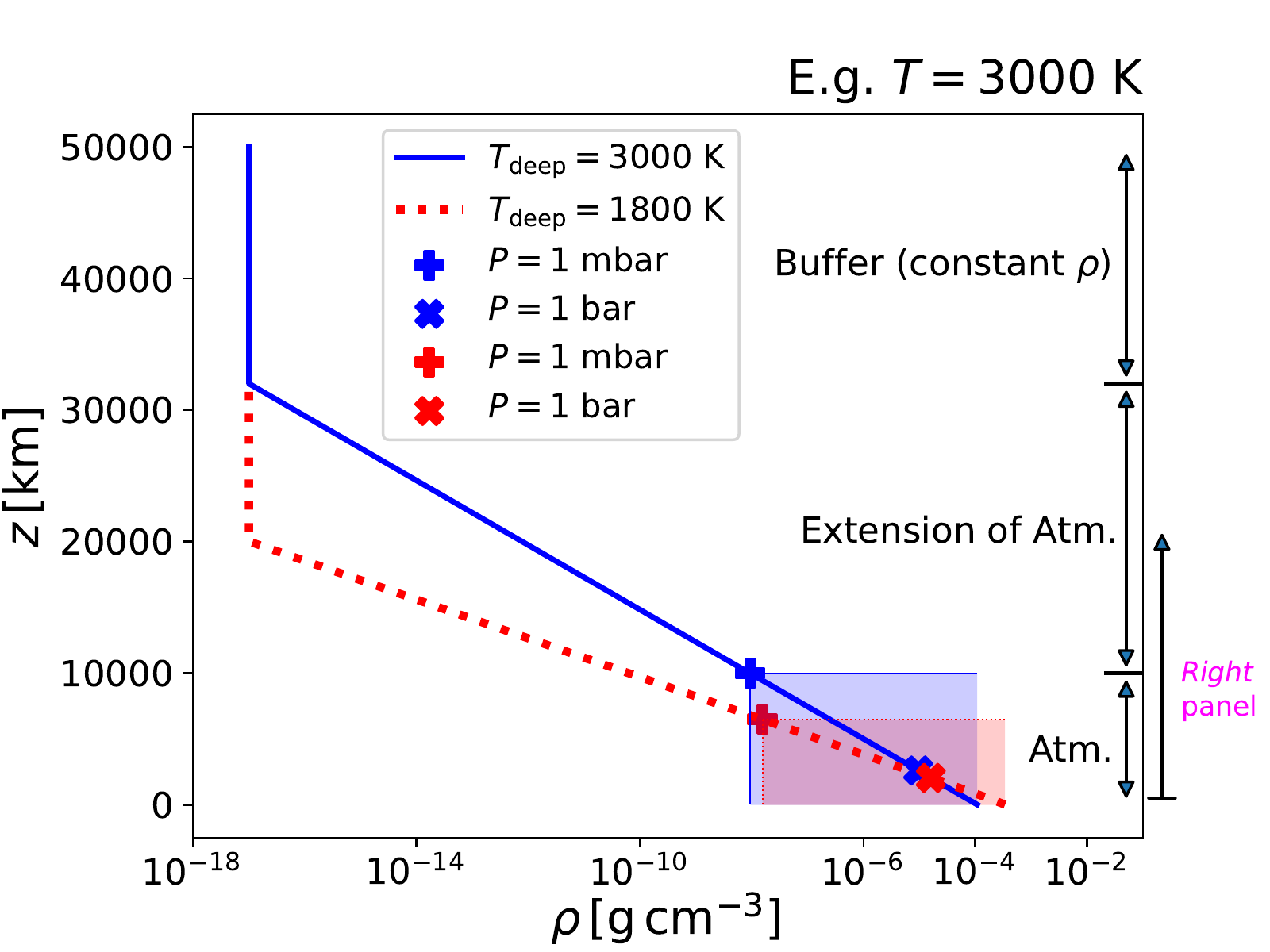}
	\includegraphics[width=8.4cm]{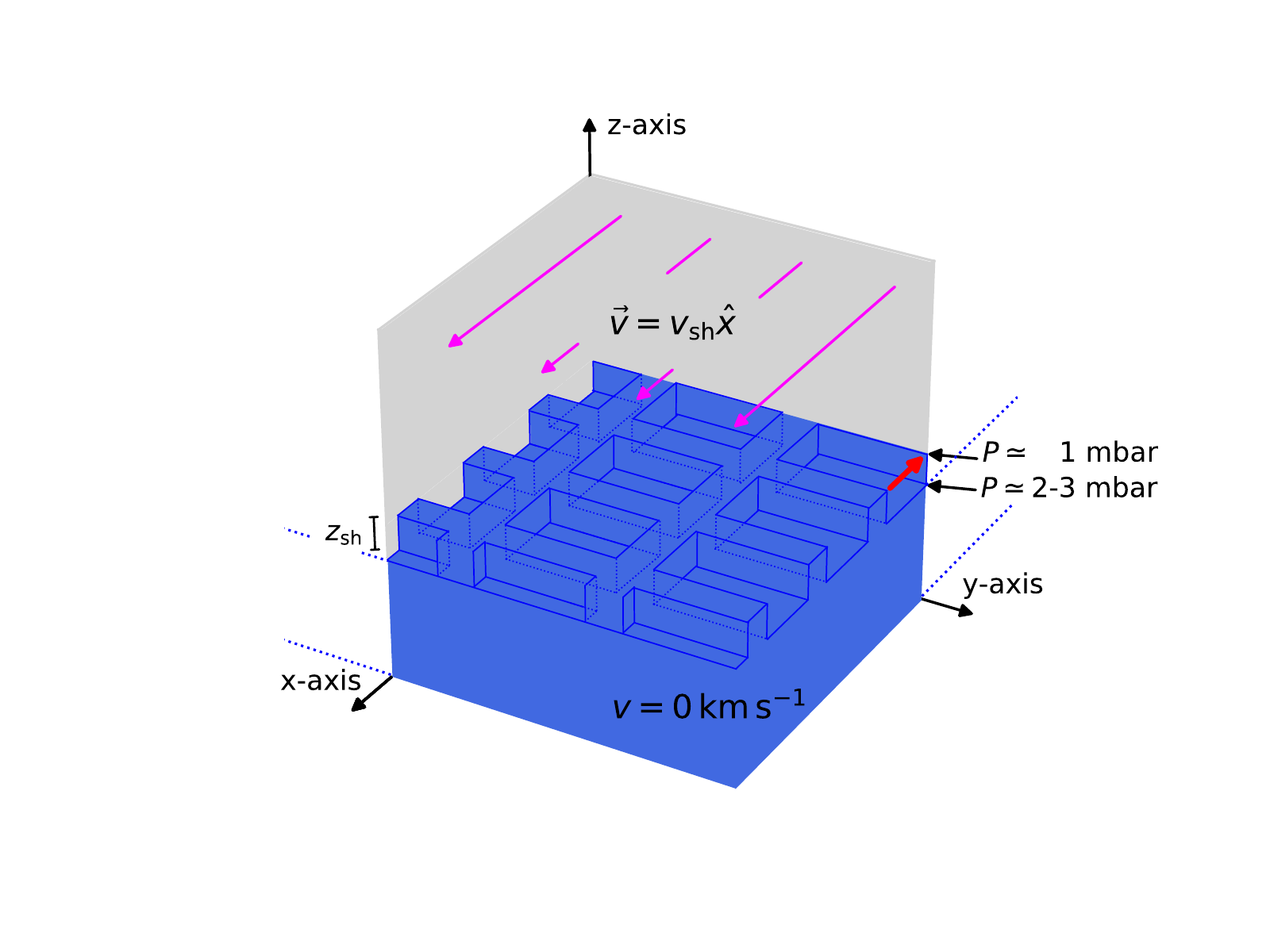}
	\caption{\textit{Left} panel: The $\rho-z$ plot for $T=1800\K$ and $3000\K$. The shaded regions on the bottom-right corner indicate the atmospheres with $P\ge1\Pmbar$. In particular, the annotations near the right vertical axis (i.e. Buffer, Extension of Atm. and Atm.) refer to the case $T=3000\K$. Furthermore, the arrow outside the same vertical axis, annotated with "\textit{Right} panel" in magenta, roughly shows the spatial scale of the schematic diagram of the atmosphere in our computation box, which is shown in the \textit{right} panel. \textit{Right} panel:  Schematic diagram of one corner of the computation box for our fiducial model. The blue region indicates our model atmosphere with initially $v=0\km\s^{-1}$. On top of it, the gas is given a shear velocity $v=v_{\rm sh}$ in the $x$ direction, indicated with magenta arrows. A shear layer in between, extending from $P=1\Pmbar$ to $P\simeq 2-3\Pmbar$, has a positive vertical gradient of $v_{\rm x}$. Notice that the relative size of the corrugations shown in this plot is exaggerated for visualization.}
	\label{fig:wiggle}
\end{figure*}

\begin{table*}
	\centering
	\setlength\extrarowheight{4pt}
	\caption{Model parameters considered in this study. We categorize the parameters into two groups: model parameters that all models  share, and those which differ among models. From left to right: The common parameters include the number of cells $N_{\rm cell}$, the number of cells per scale height $H/\Delta l$ ($H$: scale height), the gravitational constant $g$ [cm s$^{-2}$], and  the shear velocity in units of the Mach number $\mathcal{M}_{\rm sh}$. In the category of the different parameters, we list the size of our computation box $L~[10^{4}\km]$, the height $z_{\rm P=1\Pmbar}$ $[10^{4}\km]$ at $P=1\Pmbar$, the initial $T$ in the radiative zone $T_{\rm deep}$ [K], $P$ at the radiative-convective boundary $P_{\rm RCB}$ [bar], $P$ at the bottom of the domain $P_{z=0}$ [bar], the Richardson number $Ri$, estimated at $t=0$ and the sound speed $c_{\rm s}$ [km s$^{-1}$].}
	\label{table:modelparameter}
	\begin{tabulary}{1\linewidth}{c| c c c c | c c c c c  c c}
		\hline
		\multirow{2}{*}{Model name} &  \multicolumn{4}{c|}{same model parameters} & \multicolumn{7}{c}{different model parameters}\\
		\cline{2-12}
		& $N_{\rm cell}=(N_{x},~N_{y},~N_{z})$	& $H/\Delta l$  	 & $g$   & $\mathcal{M}_{\rm sh}$ & $(L_{x},~ L_{y},~ L_{z})$      & $z_{\rm P=1\Pmbar} $   & $T_{\rm deep}$ & $P_{\rm RCB}$ & $P_{z=0}$& $Ri (t=0)$ & $c_{\rm s}$\\
		\hline
		$T3000-Ri0.02$& \multirow{6}{*}{(512, 512, 1024)}  	&   \multirow{6}{*}{20}     &\multirow{6}{*}{$10^{3}$}  & \multirow{6}{*}{1} 	&	\multirow{3}{*}{$(2.5,~2.5,~5.0)$}	& \multirow{3}{*}{$1.00$}&  \multirow{3}{*}{3000} & \multirow{3}{*}{268} & \multirow{3}{*}{12}& 0.02 &\multirow{3}{*}{$3.86$}\\
		$T3000-Ri0.1~~$	&   &	 &  &	&	&  &  & & & 0.1 & \\
		$T3000-Ri0.25$		&    &	 &  &	&	&  & & & & 0.25 & \\
		$T1800-Ri0.02$		&   &	 &  &	& \multirow{3}{*}{$(1.6,~1.6,~3.2)$}	&  \multirow{3}{*}{$0.64$}& \multirow{3}{*}{1800} & \multirow{3}{*}{44} & \multirow{3}{*}{23}& 0.02 & \multirow{3}{*}{$2.99$} \\	
		$T1800-Ri0.1~~$		&   &	 &  &	& 	&  &  &  & & 0.1 &  \\	
		$T1800-Ri0.25$		&   &	 &  &	& 	& &  &  & & 0.25 &  \\	
		\hline
		\multicolumn{11}{c}{The same six models above, but with a lower resolution $N_{\rm cell}=(256,256,512)$}\\
		\hline
	\end{tabulary}
\end{table*}

\vspace{0.2in}

\subsection{Model description}
\label{sec:modeldescription}

In order to follow the evolution of our model atmosphere, we solve the
3-dimensional hydrodynamic equations in a Cartesian coordinate system,
using the code CASTRO \citep{Almgren+2010}.  CASTRO is an
adaptive mesh, compressible radiation-hydrodynamics simulation code,
based on an Eulerian grid. It supports a general equation of state,
nuclear reaction networks, rotation, and full self-gravity. The fully
compressible equations computed in the code CASTRO are as follows,

\begin{align}
\frac{\partial\rho}{\partial t}&=-\nabla \cdot (\rho \mathbf{u}),\\
\label{eq:hydro2}
\frac{\partial(\rho \mathbf{u})}{\partial t}&=~~\nabla\cdot (\rho\mathbf{u}\mathbf{u})-\nabla P+\rho\mathbf{g}+\textbf{S}_{\rm src},\\
\label{eq:hydro3}
\frac{\partial(\rho E)}{\partial t}&=-\nabla\cdot(\rho \mathbf{u}E+P\mathbf{u})+\rho \mathbf{u}\cdot\mathbf{g}+\mathbf{u}\cdot\textbf{S}_{\rm src},
\end{align}

\noindent where $\rho$, $\mathbf{u}$, and $P$ are the density, velocity vector, and pressure, respectively. $E$ represents the total specific energy, given by the sum of the internal energy $e$ and the kinetic energy, i.e., $E=e+\mathbf{u}\cdot\mathbf{u}/2$. $\textbf{S}_{\rm src}$ is a user-specified momentum source term, which will be described in more detail in \S\ref{sec:shearprescription}. CASTRO is suitable for capturing small scale structures of turbulence, which is our primary goal in this study.

\begin{figure*}
	\centering
	\includegraphics[width=8.2cm,height=5.7cm]{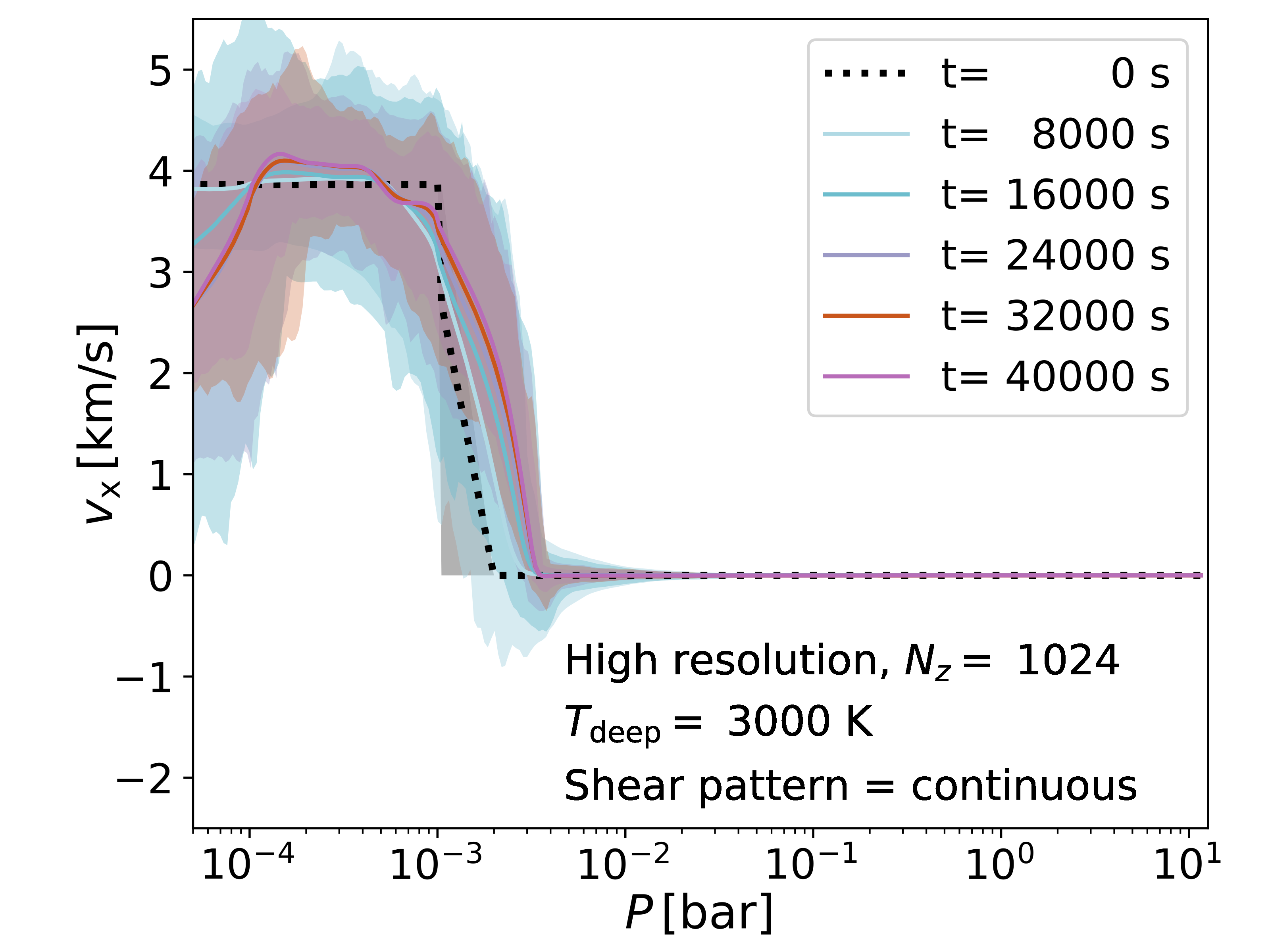}
	\includegraphics[width=8.2cm,height=5.7cm]{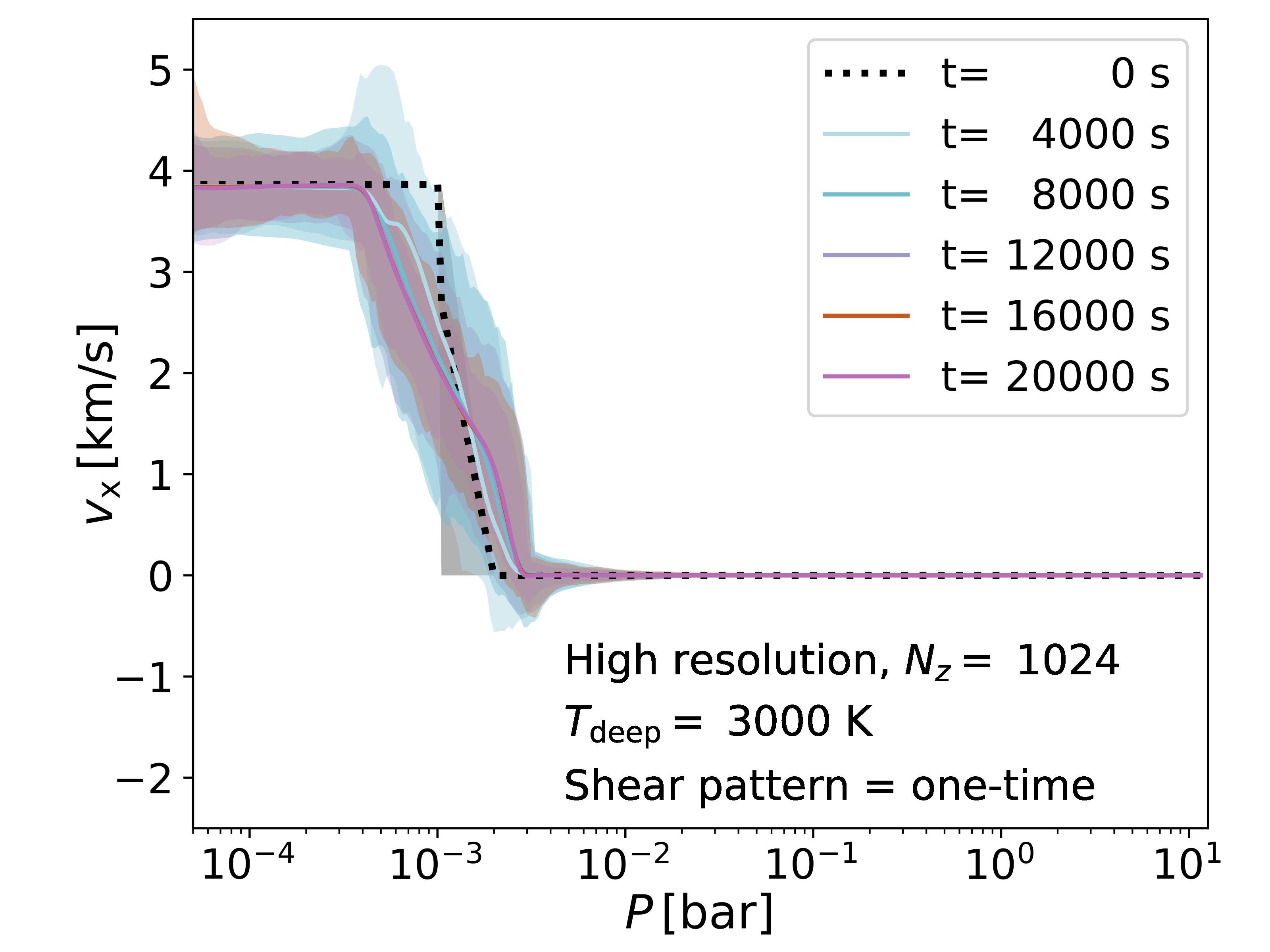}\\
	\includegraphics[width=8.2cm]{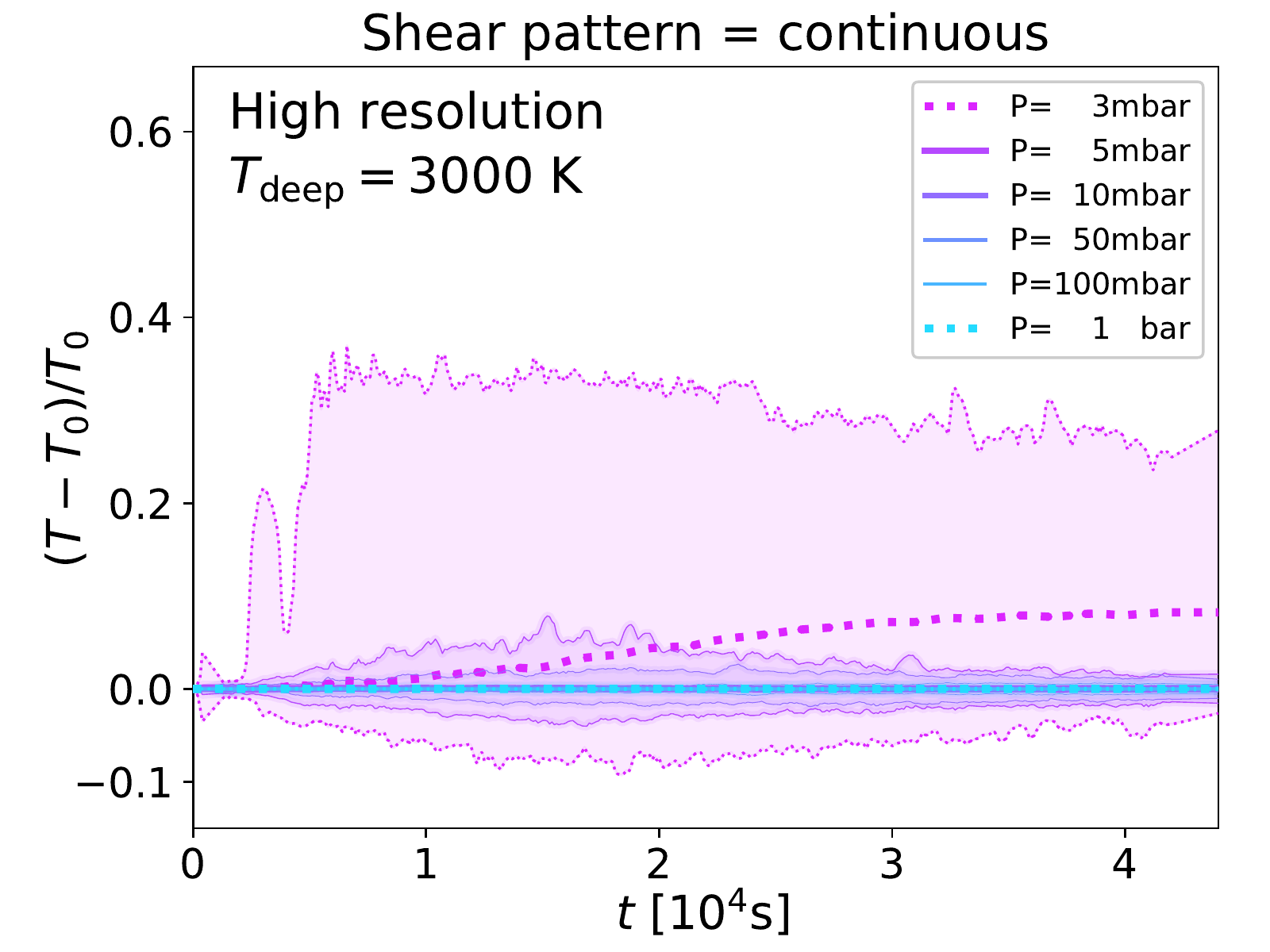}
	\includegraphics[width=8.2cm]{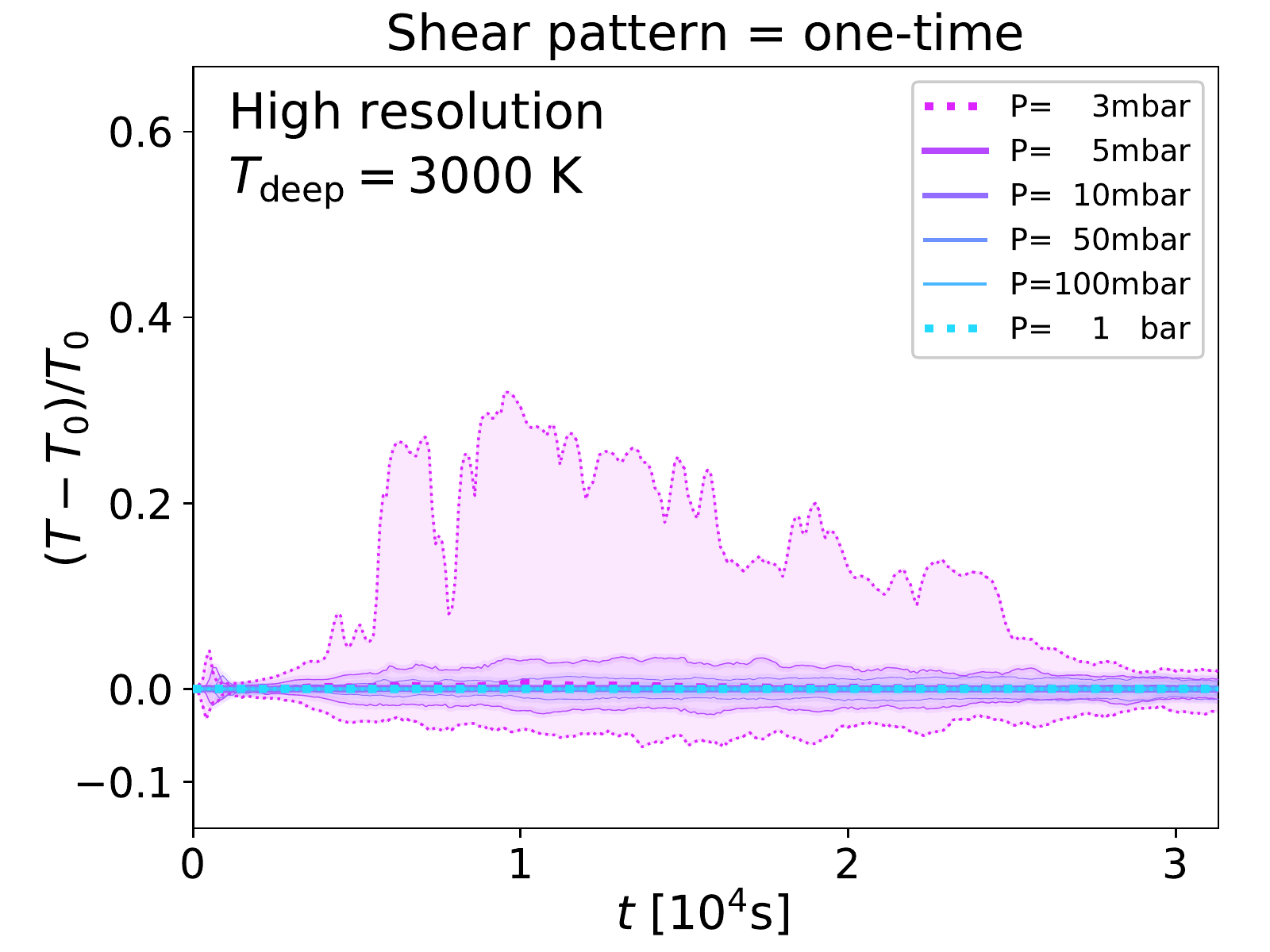}
	\caption{The velocities in the $x$ direction (\textit{top} panels) and the temperature (\textit{bottom} panels) of the atmospheres with $T_{\rm deep}=3000\K$ and $N_{\rm z}=1024$ with (``continuous'', \textit{left} panel) and without (``one-time'', \textit{right} panel) a momentum source. In the \textit{left} panel, $v_{\rm x}$ at $0.1\Pmbar\lesssim P\lesssim 1\Pmbar$  is gradually driven to be $c_{\rm s}$.  The shaded regions around the horizontal average values (solid lines) demarcate the ranges between the maximum and minimum values at a given pressure and time. Note the different timescales on the x-axes between the two panels. The simulation with the ``one-time" pattern stops at a shorter $t$ since the atmosphere becomes steady earlier.
	}
	\label{fig:vx_3000}
\end{figure*}

We consider a three-dimensional computational domain with the shape of a rectangular prism (the height is twice the width).
In our simulations, we model the radiative region of strongly-irradiated planets assuming it is initially in hydrostatic equilibrium with a constant gravity $g=10^{3}\cm\s^{-2}$. We fill the domain with our model atmosphere, starting at the bottom at $P\simeq 10\Pbar$, and approaching the top so that the atmosphere at a level of $P\simeq1\Pmbar$ occupies around 20\% of the entire domain. We define $P=1\Pmbar$ as the top of the atmospheres in this study. Then, we further extend the atmosphere until the density becomes smaller than $\rho=10^{-17}\gram \cmcube$ ($\sim44\%$ of the domain). We fill the rest of the domain ($\sim36\%$) with a constant density medium with $\rho=10^{-17}\gram \cmcube$ and $T=10^{-2}\K$. 
We refer to this region as a ``buffer''. We introduce this region to avoid possible spurious effects from an upper boundary condition (more details in the next section). The \textit{left} panel in Figure \ref{fig:wiggle} presents the $\rho-z$ plot for two different temperatures ($T=1800\K$ and $3000\K$). The annotations near the right vertical axis (i.e. Buffer, Extension of Atm. and Atm.) refer to the case of $T=3000\K$. The arrow outside the right vertical axis, annotated with ``\textit{Right} panel" in magenta, roughly shows the spatial scale of the schematic diagram of the atmosphere in our computation box shown in the \textit{right} panel. The \textit{right} panel of Figure~\ref{fig:wiggle} will be explained in detail in \S\ref{sec:shearprescription}.

To ensure robustness of our results given the numerical accuracy, we run simulations for a given initial condition with two resolutions. In the lower resolution simulations, the number of cells $N_{\rm cell}=(N_{\rm x},~N_{\rm y},~N_{\rm z})=(256,~256,~512)$, while in the higher resolution simulations, $(N_{\rm x},~N_{\rm y},~N_{\rm z})=(512,~512,~1024)$. We choose the spatial scale of each single cell to be $\sim0.1~H$ (where $H$ is the scale height) for the lower simulation case and $\sim0.05~H$ for the higher resolution. By filling the domain in this way, the total box size varies depending on what temperature we assume for the atmosphere. As will be explained in \S \ref{sec:modeldescription}, we consider two different temperatures and the total box sizes of our models are given in Table \ref{table:modelparameter}. 

Since our results are found to converge between the simulations with the two resolutions, in the result section we only focus on the atmosphere below $P\simeq1\Pmbar$ (corresponding to $z\simeq 10~H$) for further analysis in the high resolution simulations. We also discuss the differences between the simulations with the two resolutions in \S\ref{sec:T_comparison}.

\subsection{Boundary conditions}
\label{sec:boundarycondition}
We consider a different boundary condition (BC) for each boundary. We
use periodic BCs for the side boundaries. For the BC at the bottom, we
employ a ``hydrostatic" BC to provide the pressure support for
the atmosphere against gravity.  Here, the ghost cells
outside the domain are initialized to satisfy 
hydrostatic equilibrium with adjacent cells, together with 
the equation of state.  This is solved using the
Newton-Raphson method with a tolerance of $10^{-12}$.  Furthermore we
use a reflective BC on the velocity (or the momentum).  These hydrostatic
boundary conditions are described in \citet{ppm-hse}.

For the top BC, we employ an inflow boundary condition in which ghost
cells are updated to be the same as the uppermost inner cells, except
for momentum. We only allow for incoming flows (i.e. gas with a
negative vertical velocity $v_{\rm z}$). 
However, if the gas at the boundary has a positive vertical velocity $+v_{\rm z}$, it is reset to be zero.  In either case, the $x$ and $y$
components of velocities ($v_{\rm x}$ and $v_{\rm y}$) are always zero
at the boundary. The top BC is not relatively well-defined compared to
the BCs at the other sides. Hence we introduce a buffer region on top of the atmosphere to place the atmosphere of our interest sufficiently far
away from the upper boundary of the domain. This may increase the
computational cost as the volume of the entire domain (low $P$
atmosphere+buffer region) grows. However, this, along with the sponging applied in the buffer layer (see \S\ref{sec:shearprescription}), ensures that the actual top BC does not matter and our results are robust against different choices for the upper BCs.

Employing these BCs, we have confirmed that our atmospheres stay in hydrostatic equilibrium (with no forced turbulence) sufficiently longer ($t>1.5\times10^{5}\s$ for low resolution simulations and $t>3.0\times10^{5}\s$ for high resolution simulations), than the total physical times considered in this paper, namely, $t<5\times10^{4}\s$, which corresponds to a time long enough that our model atmospheres with the largest velocity gradient have become stable. Stability occurs at $t\lesssim 25~\tau_{\rm cross}$ (see Equation \ref{eq:teddy} for the definition of $\tau_{\rm cross}$).

\subsection{Shear prescription}
\label{sec:shearprescription}

It is found in numerical studies \citep[e.g][]{Guillot2002,Showman2002} that shear motions in the atmospheres of hot Jupiters can be caused by forcing due to the day-night temperature contrast. Furthermore, a typical Mach number at a level of $P=1\Pmbar$ is found to be around $\mathcal{M}\simeq1-2$ \citep[e.g][]{Showman2002,Fromang+2016}. Motivated by these studies, we give an initial shear velocity in our model atmospheres as follows. 

We consider a bulk shear motion only in the $x$
direction.  At $t=0\s$, we give a constant shear velocity at a sound
speed $c_{\rm s}$ at $P_{\rm sh}=P\lesssim 1\Pmbar$. Below this region
(higher $P$), we place a shear layer with a positive velocity
gradient, i.e. $v_{\rm x}(z=z_{i+1})-v_{\rm x}(z=z_{i})>0$ (cell index
$i$, increasing with $z$). We refer to this region as "shear layer"
throughout the paper. The velocity at the top of the shear layer is 
set to be $c_{\rm s}$ at $P=1\Pmbar$, decreasing linearly down to
$v_{\rm x}=0$ at the bottom of the shear layer. In other words, for
two adjacent cells, $\Delta v_{\rm x}/\Delta z=c_{\rm s}/z_{\rm sh}= $
constant. Here, $z_{\rm sh}$ is the height of the shear
layer. Furthermore, we consider a small number of zero velocity
corrugations in the $x$ and $y$ directions (similarly to corrugations
usually seen in billow clouds). This is to invoke more non-regular
turbulent motions in every direction, albeit the corrugated pattern is
regular. Velocity profiles (width, size and frequency of corrugation)
inside the shear layer are unknown; future work specific to this will
be necessary for more realistic modelling inside the shear layer. A
schematic diagram illustrating the shear is shown in Figure \ref{fig:wiggle}. We
emphasize that this forcing within the shear layer is only given at $t=0$. Below
the shear layer, starting typically from $P_{\rm
  sh,bottom}\simeq2-3\Pmbar$ ($P_{\rm sh,bottom}$ refers to the
pressure of the bottom of the shear layer), the atmosphere is assumed
to be in static equilibrium with no initial velocity forcing.

At $t>0\s$, we consider a momentum source in the $x$ direction within $0.1\Pmbar<P_{\rm src}<1\Pmbar$ to continuously drive shear motions. This is to mimic the east-west stream found in many global circulation models, which is likely to last as long as the rotation of a planet is synchronized with the orbital motion. For the atmosphere within this pressure range, we add a certain amount of momentum to each cell equally at every time step such that horizontal average velocities in the $+x$ direction $\overline{\textbf{v}}_{\rm x}$ gradually approach $v_{\rm x}=c_{\rm s}$ over a certain time ($t_{\rm src}=1000\s$). While the atmosphere is dynamically evolving, it is possible that a mean motion of gas at any given time at some pressure happens to be supersonic in the $+x$ direction (i.e. $\overline{\textbf{v}}_{\rm x}>c_{\rm s}\hat{\textbf{x}}$, where $\hat{\textbf{x}}$ refers to the basis vector in the $x$ direction). Whenever that is the case, we do not apply this forcing to the gas at that pressure. This way, a bulk motion is gradually driven while small structures remain intact. We note that the lower pressure limit ($P=0.1\Pmbar$) for the momentum input is arbitrarily chosen, but the atmosphere at $P>1\Pmbar$ is not sensitive to different choices of the lower limit. We can summarize this continuous forcing within $0.1\Pmbar<P_{\rm src}<1\Pmbar$ as follows. For gas at a given pressure $P$ and time $t$ with a mean motion $\overline{v}_{\rm x}$, an external momentum (i.e. \textbf{\textit{S}}$_{\rm src}\Delta t$ in Equation~\ref{eq:hydro2}) is added to the momentum of the gas at each cell 
\begin{align}
\label{eq:forcing}
\textbf{S}_{\rm src}=\overline{\rho}\frac{\max(0,~c_{\rm s}-\overline{v}_{\rm x})}{t_{\rm src}},
\end{align}
where ``max()" indicates the maximum of the two values in the parenthesis. To conserve the total energy of the gas, we take into account its additional energy accordingly (see Equation \ref{eq:hydro3}).

As an example, Figure \ref{fig:vx_3000} shows different evolutions of gas motions in the $x$ direction and the temperature of the atmosphere with (``continuous'', \textit{left} panel) and without (``one-time'', \textit{right} panel) the additional momentum source. In the \textit{upper} panel, we show $v_{\rm x}$ throughout the atmosphere at different times and in the \textit{lower} panel the relative temperature variations with respect to the initial temperature $T_{0}$ at different pressure levels. In all the plots, the shaded regions around the horizontal average values (solid lines) demarcate the ranges between the maximum and minimum values at a given pressure and time. The size of the shaded regions is relevant for our study, along with the average values. This is because it can serve as a good indicator for how chaotic the atmosphere is due to turbulence. We can see some clear differences in both the $v_{\rm x}$ and the $T$ plots. Among those, the most noticeable difference is in the larger shaded regions with the extra momentum source. This means that the momentum source clearly contributes to amplifying the effects and lifetime of turbulent motions, as we expect. From now on, we will only consider the ``continuous'' shear case.

For $P<0.01\Pmbar$, we damp the velocity of the gas to $\textbf{v}=0$ in all directions. We use a damping scheme (or ``sponge'' damping) employed in the low Mach number code {\tt MAESTRO} \citep{Nonaka+2010} and used in other studies \citep[e.g.][]{Zingale+2009,Zingale+2011,Nonaka+2012}. This scheme was originally introduced to avoid a large growth in velocities in the low density regions of a stellar surface due to intense heating. See  Section 4.3.1 in \citet{Almgren+2008} for the equations used for the damping scheme. In our case, an unphysical surge in velocities can occur in the buffer region. This scheme serves to damp the large velocity wakes which would otherwise propagate towards the atmosphere and significantly affect its stability. 

To sum up, the atmosphere is modelled such that: \\
\noindent At $t=0$ (from $z=0$ to larger $z$),
\vspace{-0.1in}
\begin{enumerate}
	\item  $P_{\rm sh,bottom}\lesssim P$: hydrostatic equilibrium with $\textbf{v}=0$ ($P_{\rm sh,bottom}$ is determined by the size of the shear layer; see \S\ref{sec:initialcondition} below),

	\item $1\Pmbar \lesssim P\lesssim P_{\rm sh,bottom}$: positive velocity gradient in the vertical direction,
	
	\item $P\lesssim1\Pmbar$: $v_{\rm x}=c_{\rm s}$ and $v_{\rm y}=v_{\rm z}=0$.
	
\end{enumerate}
At $t>0$, the atmospheric region at $P\lesssim1\Pmbar$ is affected by:
\begin{enumerate}
	\item $0.1 \Pmbar \lesssim P\lesssim1\Pmbar$: momentum source ($v_{\rm x}\rightarrow c_{\rm s}$);
	
	\item $P\lesssim0.01\Pmbar$: sponge damping ($\textbf{v}\rightarrow0$).
	
\end{enumerate}

\begin{figure}
	\centering
	\includegraphics[width=8.2cm]{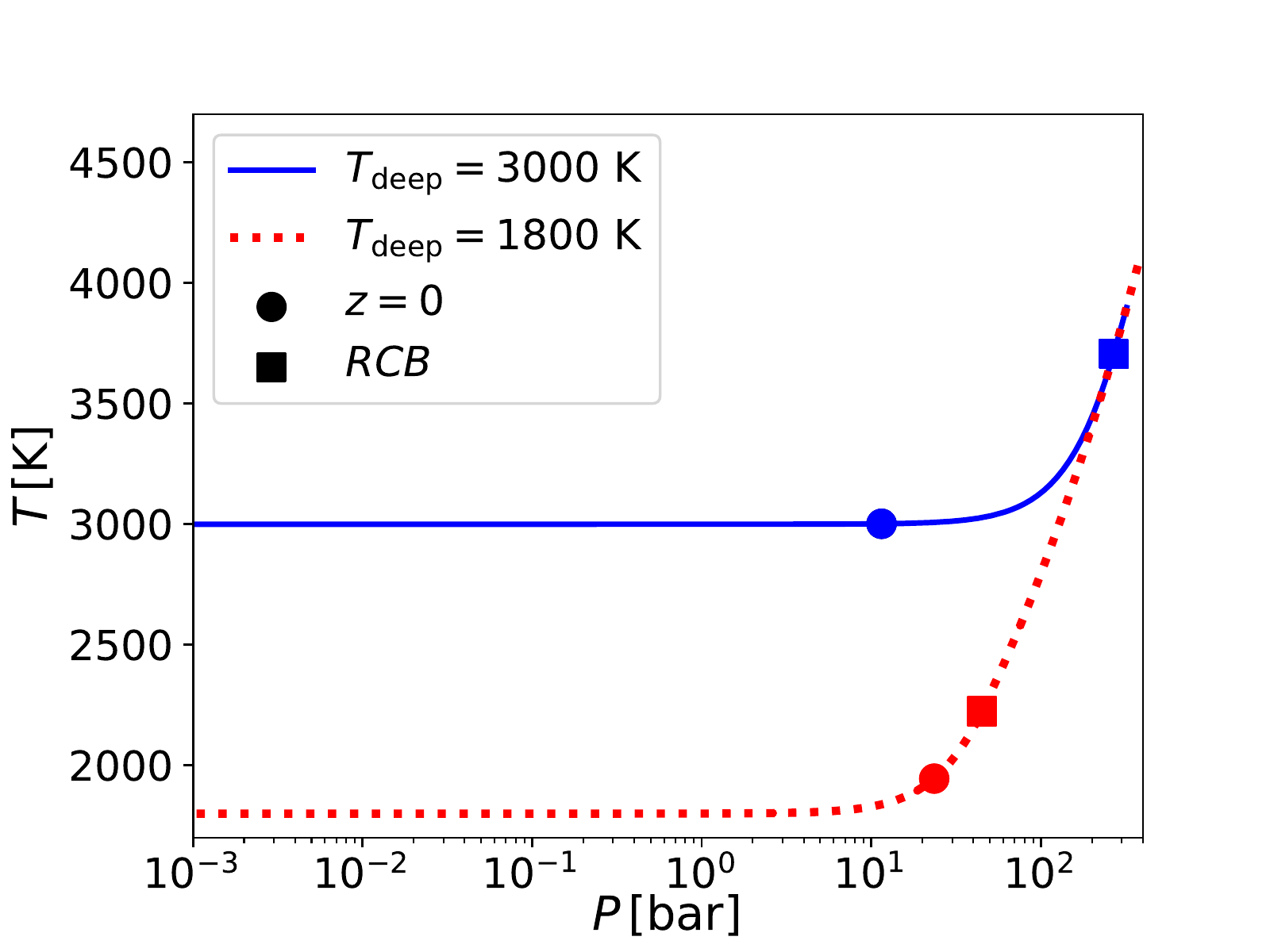}
	\caption{The initial $T-P$ profile for $T_{\rm deep}=3000\K$ (blue solid lines) and $T_{\rm deep}=1800\K$ (red dotted lines). We mark the pressure at $z=0$ with circles, sharing the same line types and colors. The squares indicate the radiative-convective boundary (RCB), estimated using Equations 12 and 13 of \citet{YoudinMitchell2010}, for the given temperatures. }
	\label{fig:initial_T_P}
\end{figure}
\subsection{Model parameters and initial condition}
\label{sec:initialcondition}

\begin{figure*}
	\hspace{-1.9in}
	\vbox to220mm{	\vspace{-0.7in} 
		\includegraphics[width=21.3cm]{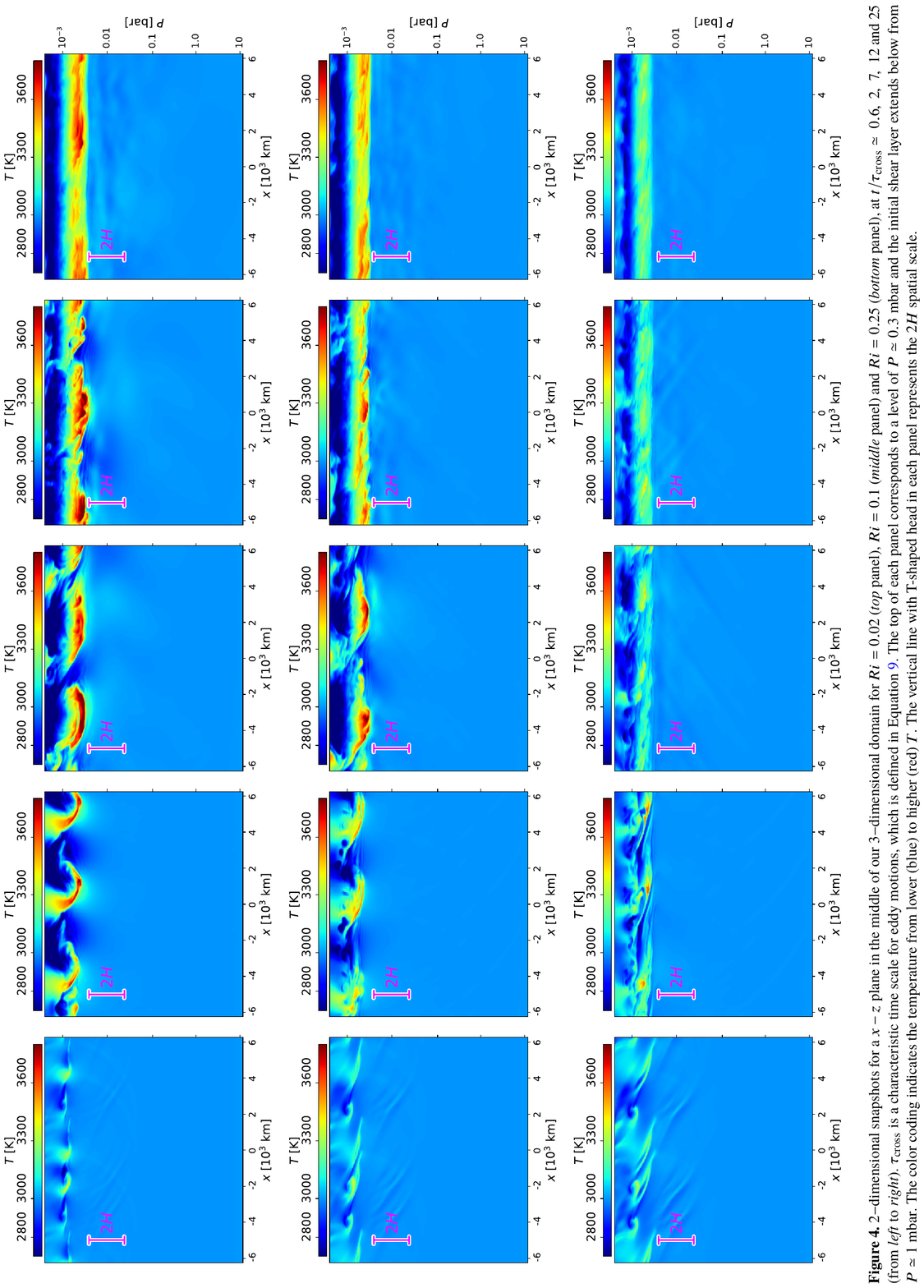}
		\caption{}
	}
	\label{fig:slice_Ri001}
\end{figure*}

Our primary goal is to capture small-scale structures of turbulence and quantify the turbulent kinetic and heat energy flux which penetrates into the atmosphere. For this, we employ the analytic model in \citet{YoudinMitchell2010} to determine the initial properties of our model atmospheres.

The atmospheres are characterized by two different temperatures, $T_{\rm deep}$ and $T_{1}$. Note that we use the same notation as in \citet{YoudinMitchell2010}. $T_{\rm deep}$ is the temperature at the top of the atmosphere. Throughout this paper, we define the top of the atmosphere to be at $P=1\Pmbar$. On the other hand, $T_{1}$ is the temperature that convective regions would have at $P=1\Pbar$, and it measures the internal entropy of the adiabat. Generally speaking, those two temperatures determine where the radiative-convective boundary (RCB) would be located. The pressure at the  RCB $P_{\rm RCB}$ is lower for larger $T_{1}$ (higher entropy) or lower $T_{\rm deep}$ (strong irradiation).

\begin{figure*}
	\centering
	\includegraphics[width=6.0cm]{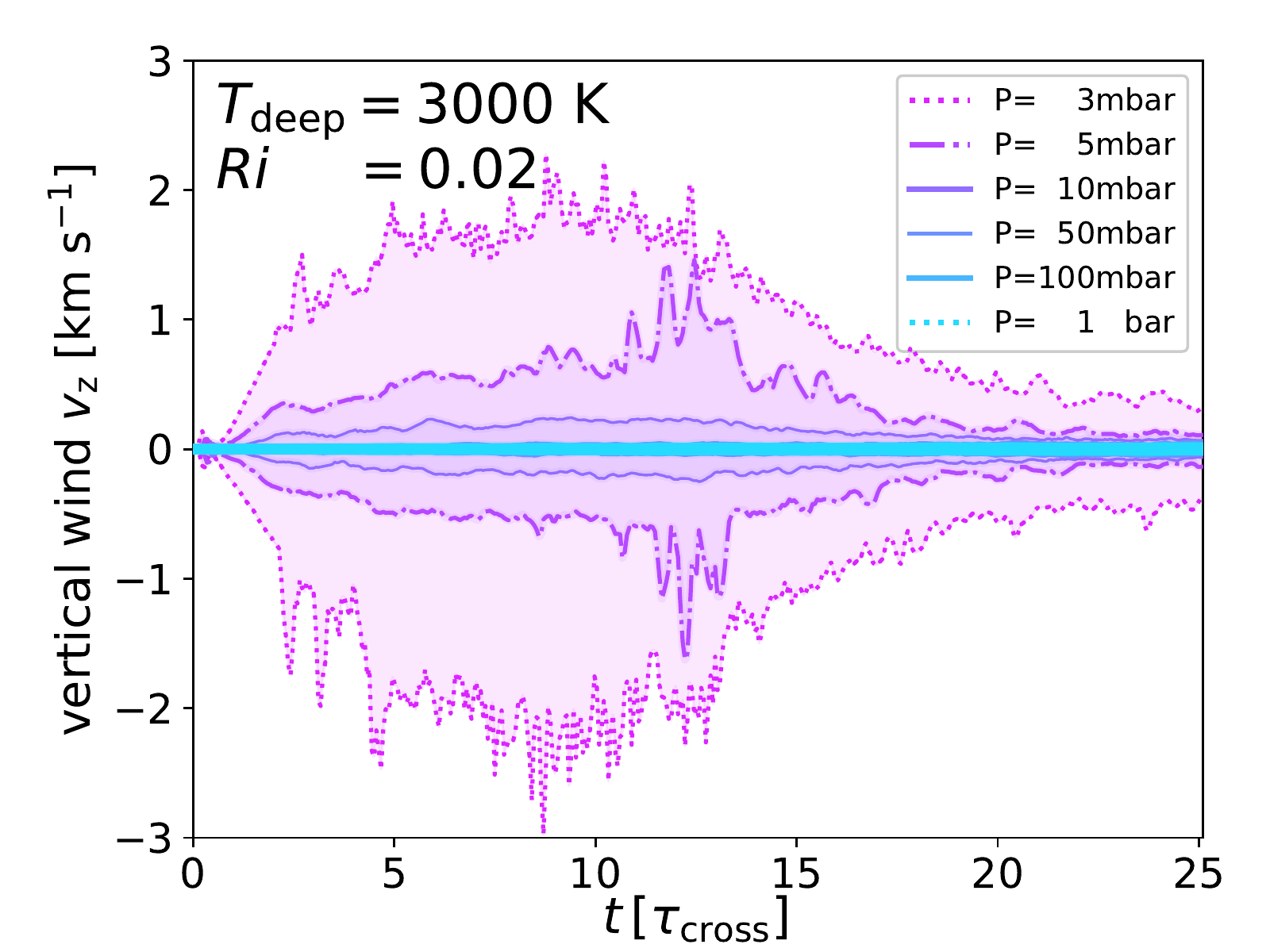}\hspace{-0.1in}
	\includegraphics[width=6.0cm]{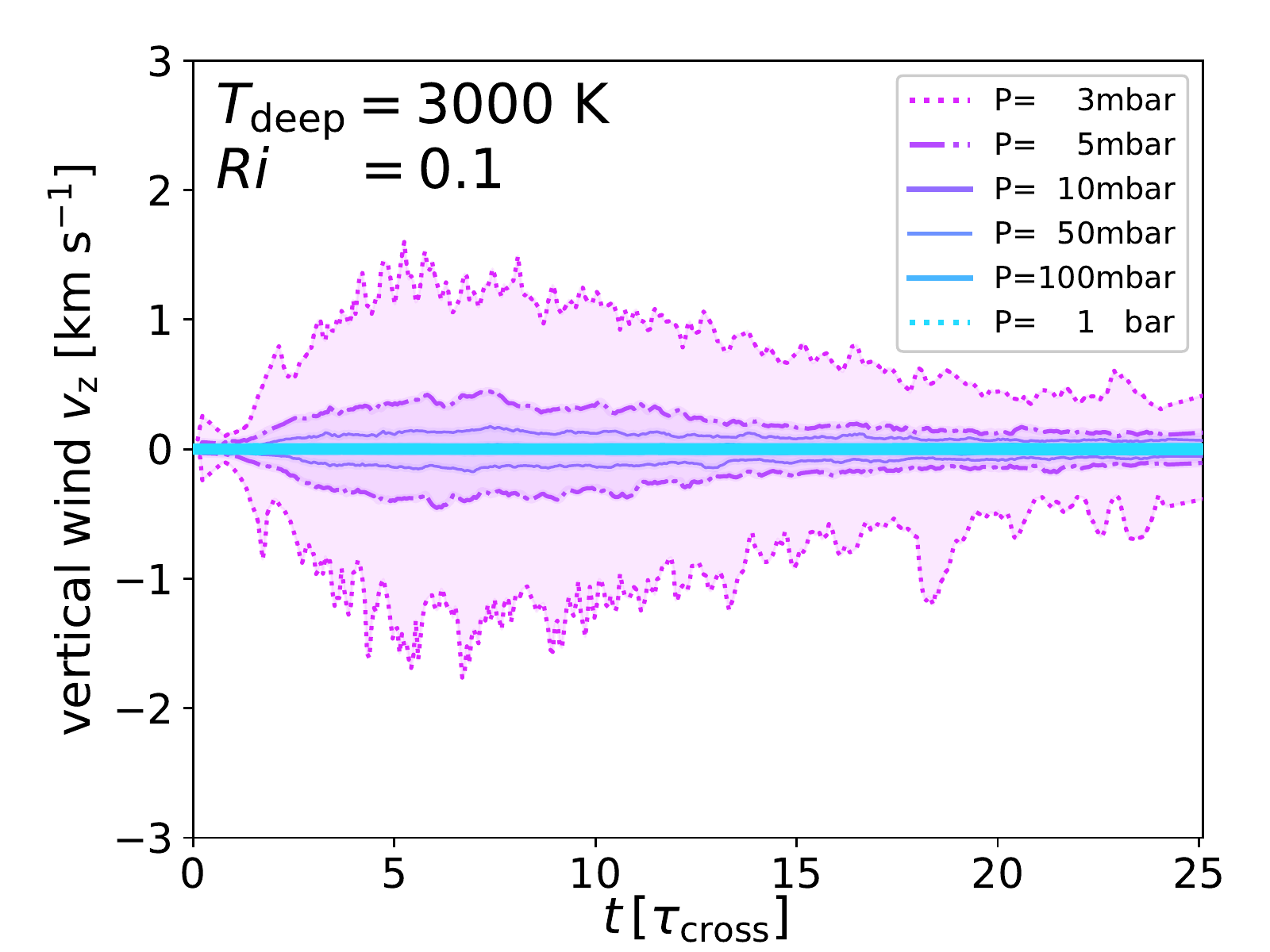}\hspace{-0.1in}
	\includegraphics[width=6.0cm]{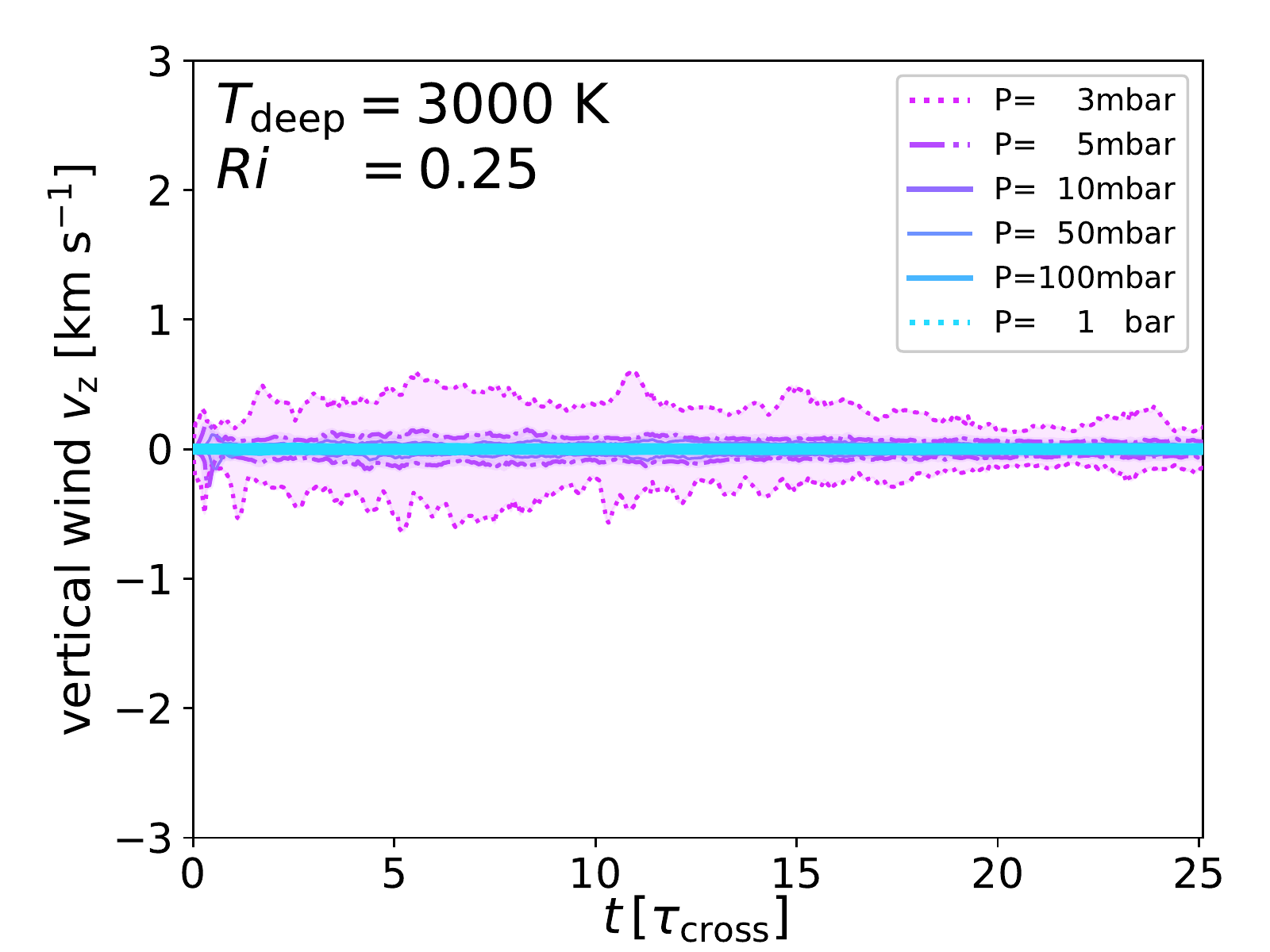}\hspace{-0.1in}
	\caption{The average vertical velocity $\overline{v}_{\rm z}$ at different pressures for (from \textit{left} to \textit{right}) $Ri=0.02,~0.1$ and $0.25$. The boundaries of the shaded regions show the maximum and the minimum values around the average values (solid lines) at a given time and pressure. }
	\label{fig:time_verticalwind}
\end{figure*}
\begin{figure*}
	\centering
	\includegraphics[width=6.0cm]{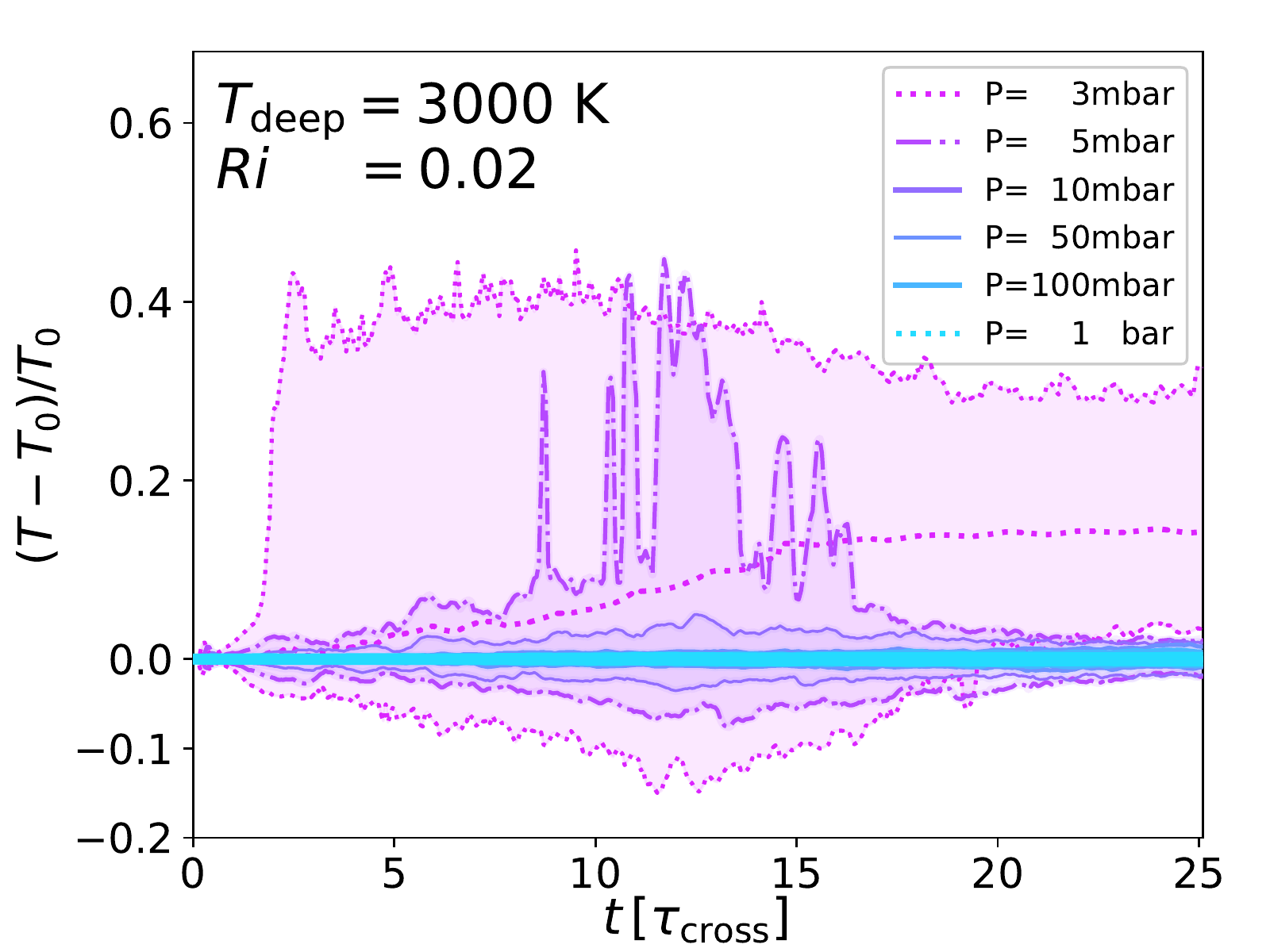}\hspace{-0.1in}
	\includegraphics[width=6.0cm]{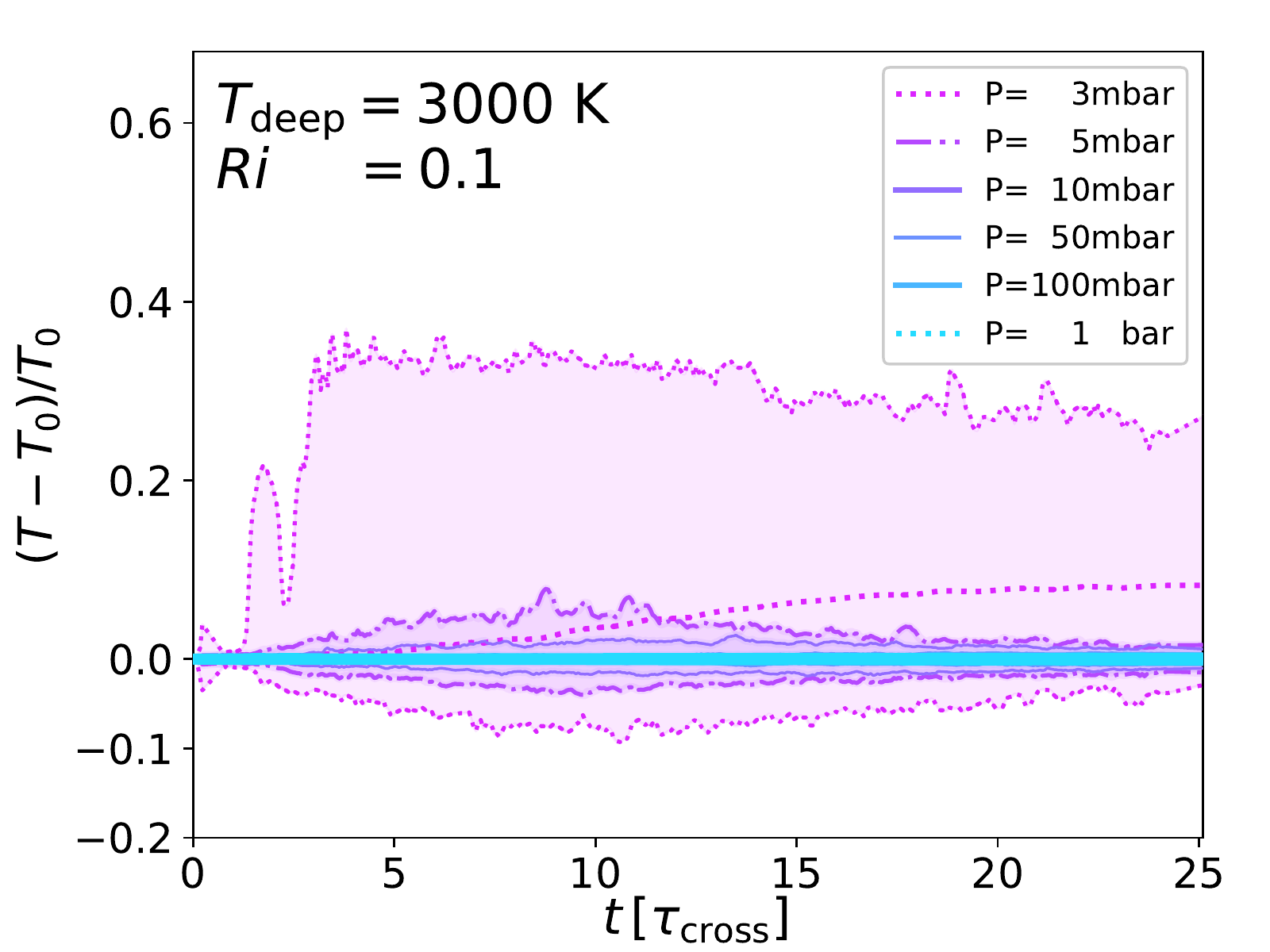}\hspace{-0.1in}
	\includegraphics[width=6.0cm]{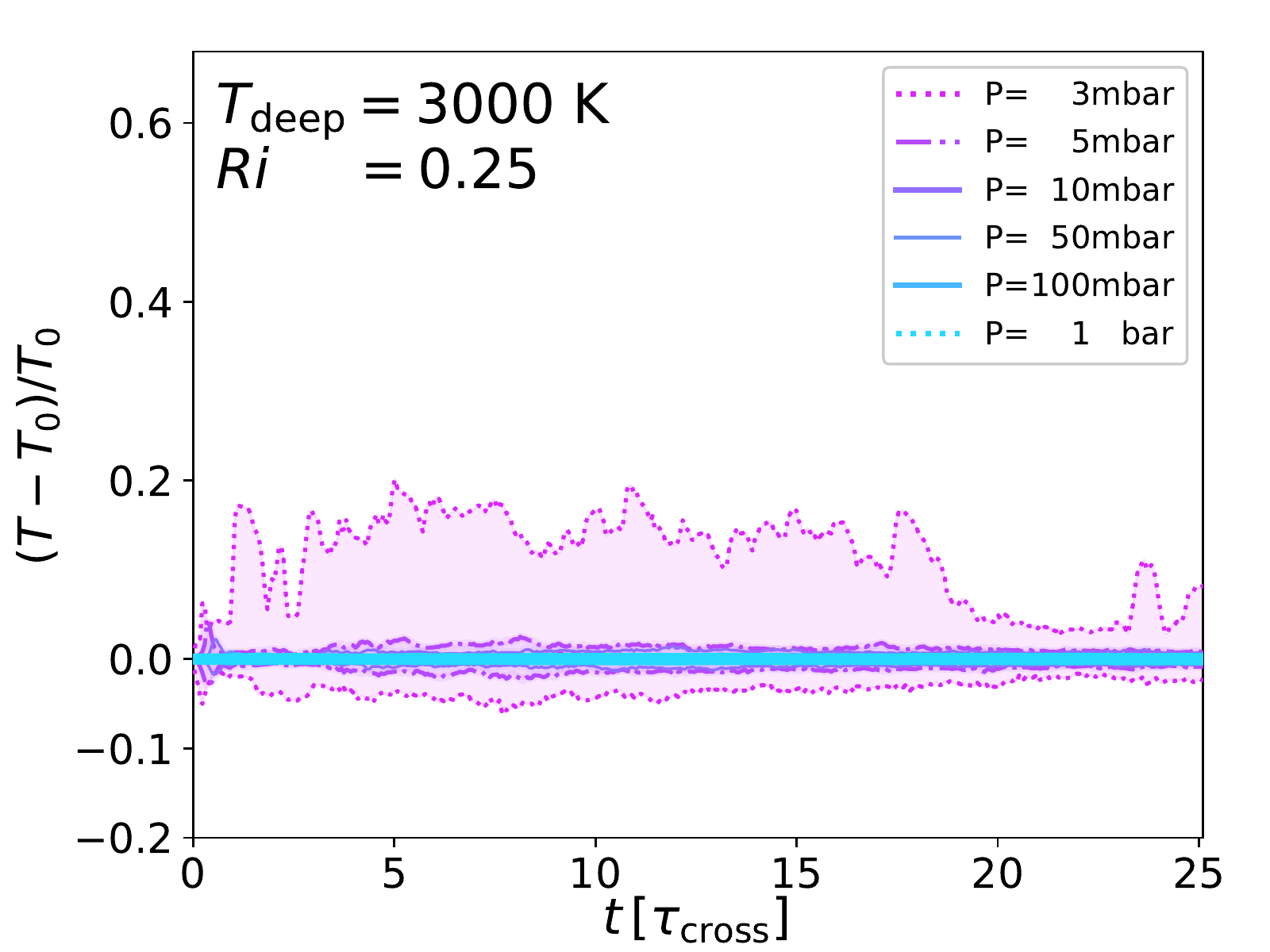}\hspace{-0.1in}
	\caption{Fractional changes of $T$ relative to the temperature at $t=0$ (denoted by $T_{0}$) for the same pressures as in Figure \ref{fig:time_verticalwind}. The shaded regions have the same meaning as before. }
	\label{fig:time_T_diff}
\end{figure*}

In this study, our fiducial model assumes $T_{\rm deep}=3000\K$ and $T_{1}=750\K$ with $P\simeq12\Pbar$ at $z=0$. In addition, we explore a case with a lower temperature with $T_{\rm deep}=1800\K$ for the same adiabat\footnote{For those two cases, $P_{\rm RCB}=268\Pbar$ and $T_{\rm RCB}=3707\K$ for $T_{\rm deep}=3000\K$ and $P_{\rm RCB}=44\Pbar$ and $T_{\rm RCB}=2224\K$ for $T_{\rm deep}=1800\K$.}. In Figure \ref{fig:initial_T_P}, we show the initial $T-P$ profiles of our models for $T_{\rm deep}=3000\K$ (blue solid lines) and $T_{\rm deep}=1800\K$ (red dotted lines), respectively. The circular marks represent the pressure at the bottom of our domain and the squares the radiative-convective boundaries (RCB) for the given temperatures.

We assume the atmosphere is made up of an ideal diatomic gas, with a mean molecular weight $\mu=2.34$, following the equation of state, 
\begin{align}
P=\rho R T=(\gamma-1)\rho e,
\end{align}
where $\gamma=7/5$ and the gas constant $R=3.6\times10^{7}\erg\gram^{-1}\cm^{-1}$. Again, $e$ is the internal energy of the gas.

One important parameter left to decide is the height of the shear layer $z_{\rm sh}$ (see Figure \ref{fig:wiggle}) or $P_{\rm sh,bottom}$, which determines the initial vertical gradient. In our simulations, eddy motions due to a Kelvin-Helmholtz instability are triggered by the non-zero velocity gradient in the shear layer ($v_{\rm x}=c_{\rm s}$ at $P\simeq 1\Pmbar$ and $v_{\rm x}=0$ at $P\simeq P_{\rm sh,bottom}$). For an incompressible flow, this is possible when the Richardson number is smaller than 1/4 \citep{Chandrasekhar1961}. The Richardson number $Ri$ is defined as follows, 
\begin{align}
Ri=\frac{N^{2}}{(dv/dz)^{2}},
\label{eq:Ri}
\end{align}
where $N_{\rm BV}$ represents the Brunt-V\"{a}is\"{a}l\"{a} frequency,
\begin{align}
N_{\rm BV}^{2}=\frac{\rho g^{2}}{P}[\nabla_{\rm ad}-\nabla].
\label{eq:BVfrequency}
\end{align}
In the above, $\nabla$ refers to the lapse rate of the atmosphere, defined as,
\begin{align}
\nabla=\left(\frac{d\ln T}{d\ln P}\right)
\label{eq:lapserate}
\end{align}
and the adiabatic lapse rate $\nabla_{\rm ad}=2/7$. 
From this expression, we can expect that if the vertical velocity gradient is chosen to be smaller, we start with a more unstable atmosphere. 
We choose the height of the shear layer to be small enough to give $Ri\lesssim0.25$. In particular, we assume $Ri\simeq0.02,~0.1$ and $0.25$. These correspond to  $N_{\rm{sh},z} = 6$, $14$ and $24$ within the layer for the higher resolution (hence half the cell number for the lower resolution case). 

Each of our models is integrated until the atmosphere reaches a steady state. We assume the atmosphere becomes steady when variables including $T$ and $\textbf{v}$ below $P\simeq 3\Pmbar$ do not change significantly. Typically, the atmospheres reach a steady state at $t\geq 10^{4}\s$ for $Ri=0.02$, the time being shorter with larger $Ri$. Note that this is still sufficiently shorter than the time for our model atmosphere to remain in equilibrium when there is no initial shear motion.

The model parameters and their initial values are summarized in Table \ref{table:modelparameter}.

\section{Results}
\label{sec:results}

In this section, we analyze the evolution of the thermodynamic properties of our model atmospheres. In particular, we focus on how much and how deep heat and kinetic energy fluxes can penetrate into the atmospheres. Additionally, we discuss shock formation in the atmospheres.

\subsection{Development of Eddies}

In all of our models, turbulent motions are created first inside the shear layer. The unstable motions spread downwards over time, but they are limited within $z\sim (1-2)~H$, as shown in Figure {\color{blue}4}. Each panel shows a $2-$dimensional snapshot of a $x-z$ plane in the middle of our $3-$dimensional box, for $Ri=0.02$ (\textit{top} panel), $Ri=0.1$ (\textit{middle} panel) and  $Ri=0.25$ (\textit{bottom} panel) at $t/\tau_{\rm cross}\simeq~0.6, ~2,~ 7,~ 12$ and  $25$ (from {left} to {right}). Here we use the crossing time of a sound wave across a pressure scale height as a characteristic time scale $\tau_{\rm cross}$, which we define as follows,

\begin{align}
\tau_{\rm cross}\simeq 2\uppi  \frac{H}{c_{\rm s}}=\begin{cases}
~~1734\s\,,\hspace{0.8in} \text{$T_{\rm deep}=3000\K$},\\
~~1343\s\,,\hspace{0.8in} \text{$T_{\rm deep}=1800\K$}.
\end{cases}
\label{eq:teddy}
\end{align}

The top corresponds to a level of $P\simeq 0.3\Pmbar$ and the initial shear layer extends below from $P\simeq1\Pmbar$. The plots are color-coded according to the temperature from blue (lower $T$) to red (higher $T$). The vertical line with T-shaped heads in each panel indicates the $2H$ spatial scale.
The temperatures in the atmospheres with lower $Ri$ are generally higher at a given time. 
Furthermore, eddies at larger scales break up into smaller eddies. These are typically expected in a standard picture of turbulence. It is noticeable that distinctive variations in $T$ are limited within a vertical range of $1-2~H$, i.e., $P\lesssim10\Pmbar$. In our models, $P\simeq2.5\Pmbar$ at $z=z_{\rm 1\Pmbar}-H$, where $z_{\rm 1\Pmbar}$ is the height at $P=1\Pmbar$,  and $P\simeq6\Pmbar$ at $z=z_{\rm 1\Pmbar}-2H$. In our model we do not consider convective bulk motions in the vertical direction. That means that heat energy transfer can only occur via turbulent motions of the gas. In light of this, we can see from the slice plots that the atmosphere below $P\simeq10\Pmbar$ is not significantly affected by turbulent motions driven in the shear layer near its top.  Note that the temperature evolution shown in the plot is not only directly indicative of heat fluxes, but it also gives us a sense of the overall effect of turbulence. This will be shown more clearly in the following sections.

\subsection{Velocity and temperature variation}

It has been suggested that turbulent mixing due to the non-linear Kelvin$-$Helmholtz instability plays an important role in the penetration and dissipation of kinetic energy. Vertical motions of gas are a primary factor to determine in which direction the kinetic and heat energy fluxes (along with temperature variations) propagate. Therefore, it is important to understand first how vertical velocities in our atmospheres evolve and how temperatures vary over time under the presence of the shear motion.

Figure \ref{fig:time_verticalwind} shows the average vertical velocity $\overline{v}_{\rm z}$ at different pressures for (from {left} to {right}) $Ri=0.02,~0.1$ and $0.25$. 
Positive (negative) values indicate upward (downward) movements. Note that $P\simeq3\Pmbar$ corresponds to the bottom of the shear layer for $Ri=0.25$, whereas for lower $Ri$ the shear layers are positioned at $P\lesssim3\Pmbar$. The shaded regions are bounded by the maximum and the minimum values around the average values (solid lines) at a given time and pressure. Shaded regions in other plots of this type below will have the same meaning. A general trend is that the magnitude of fluctuations in $v_{\rm z}$ ($|v^{'}_{\rm z}|=|\overline{v}_{\rm z}-v_{\rm z}|$) increases up to $t/\tau_{\rm cross}\simeq5-10$, but gradually decreases afterwards. The velocity fields symmetrically fluctuate and |$v^{'}_{\rm z}|$ decreases by a factor of 3--5 from one pressure to the next higher pressure chosen in the plots. Meanwhile, average vertical motions remain almost zero at those pressure levels. This means that there is no dominant bulk motion in the vertical direction and eddies are confined within a small volume (a few $\Pmbar$ vertically). This is also shown in Figure {\color{blue}4}. At $t/\tau_{\rm cross}\gtrsim10$, $|v^{'}_{\rm z}|$ does not increase, but rather decays or stays constant.

We next present the relative changes of $T$ with respect to the temperature at $t=0$ (denoted by $T_{0}$) in Figure \ref{fig:time_T_diff} for the same pressures as in Figure \ref{fig:time_verticalwind}. Somewhat contrary to the symmetric changes in $\overline{v}_{\rm z}$, as $Ri$ decreases (\textit{left} and \textit{middle} panel), turbulence leads to a more positive temperature fluctuation $T'$ at $P\simeq3\Pmbar$ (corresponding to the bottom of, or slightly below, the shear layer). However this is not clearly seen for the models with $Ri=0.25$ (\textit{right} panel) and at $P\gtrsim 5\Pmbar$ for $Ri=0.02$ and $0.1$.

\begin{figure*}
	\centering
	\includegraphics[width=5.95cm]{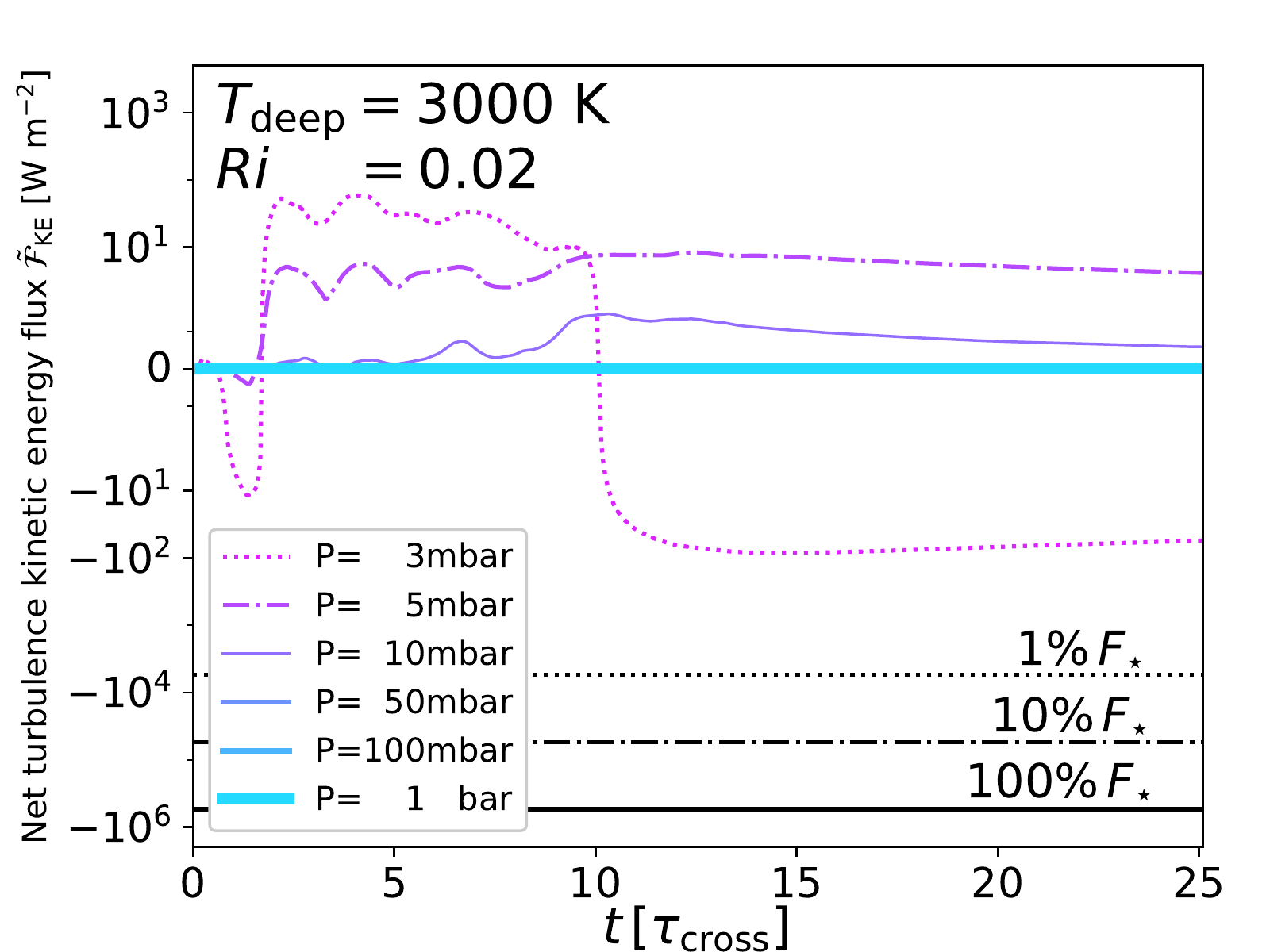}\hspace{-0.08in}
	\includegraphics[width=5.95cm]{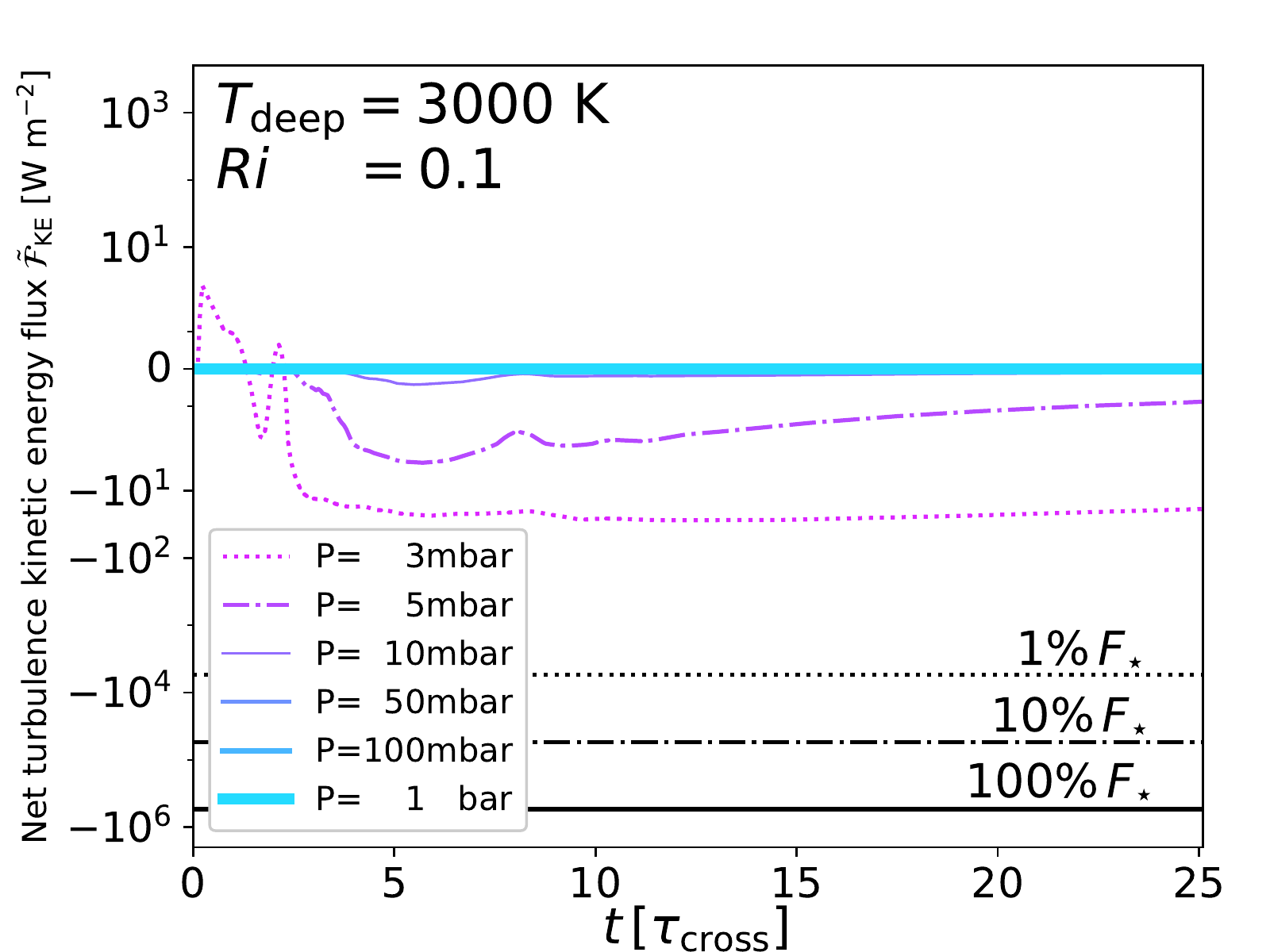}\hspace{-0.08in}
	\includegraphics[width=5.95cm]{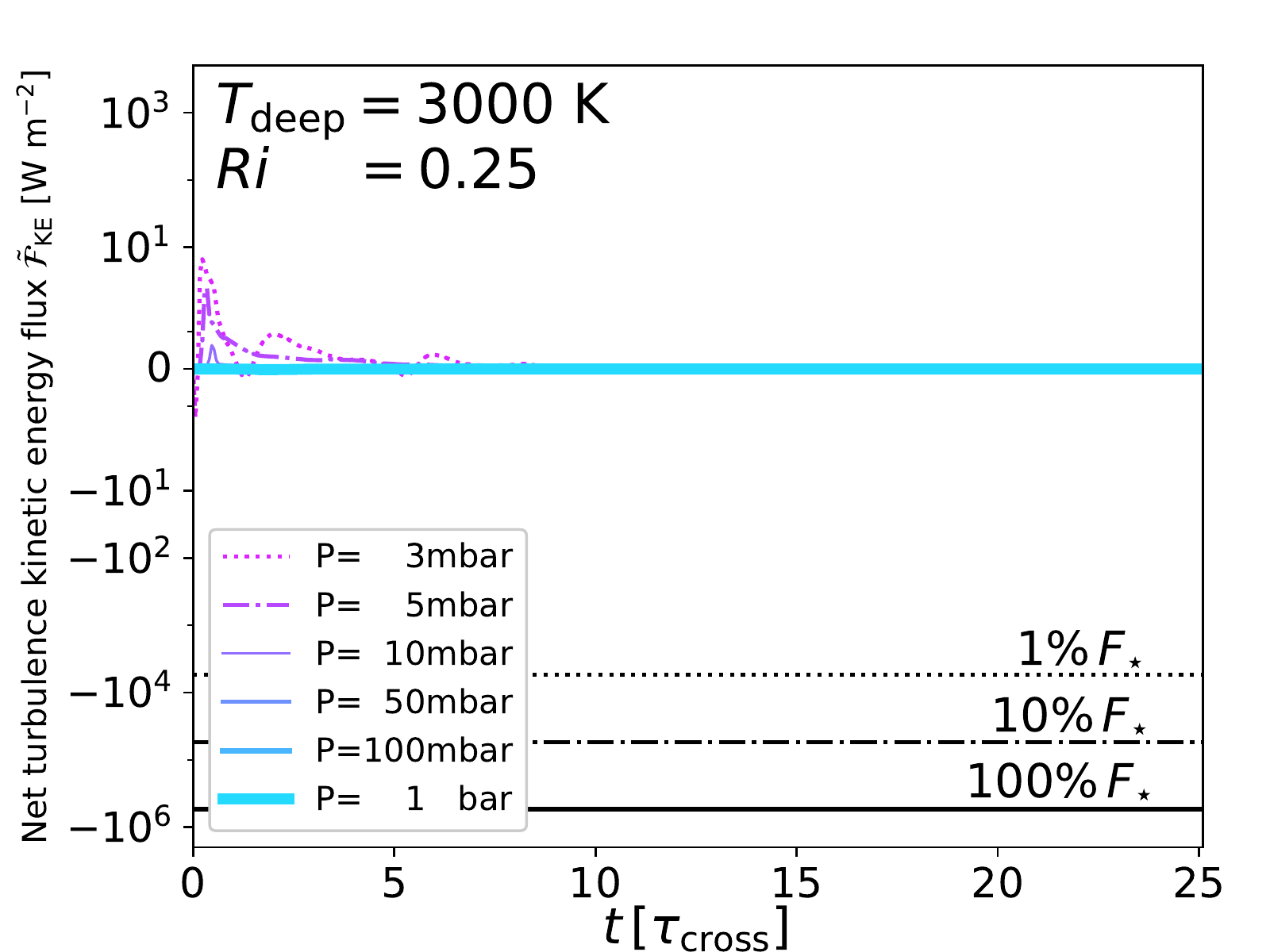}\hspace{-0.08in}
	\caption{Time evolution of the the turbulence kinetic energy flux $\mathcal{\tilde{F}}_{\rm KE}$ for $Ri=0.02$ (\textit{left} panel), $Ri=0.1$ (\textit{middle} panel) and $Ri=0.25$ (\textit{right} panel). The horizontal lines near bottom indicate 1\% ($\simeq-10^{6}~\rm{W}~\rm{m}^{-2}$, solid horizontal line), 10\% ($\simeq-10^{5}~\rm{W}~\rm{m}^{-2}$, dot-dashed line) and 100\% ($\simeq-10^{4}~\rm{W}~\rm{m}^{-2}$, dotted line) of an incoming stellar flux at $T=3000\K$. }
	\label{fig:time_turbkin}
\end{figure*}

\begin{figure*}
	\centering
	\includegraphics[width=5.95cm]{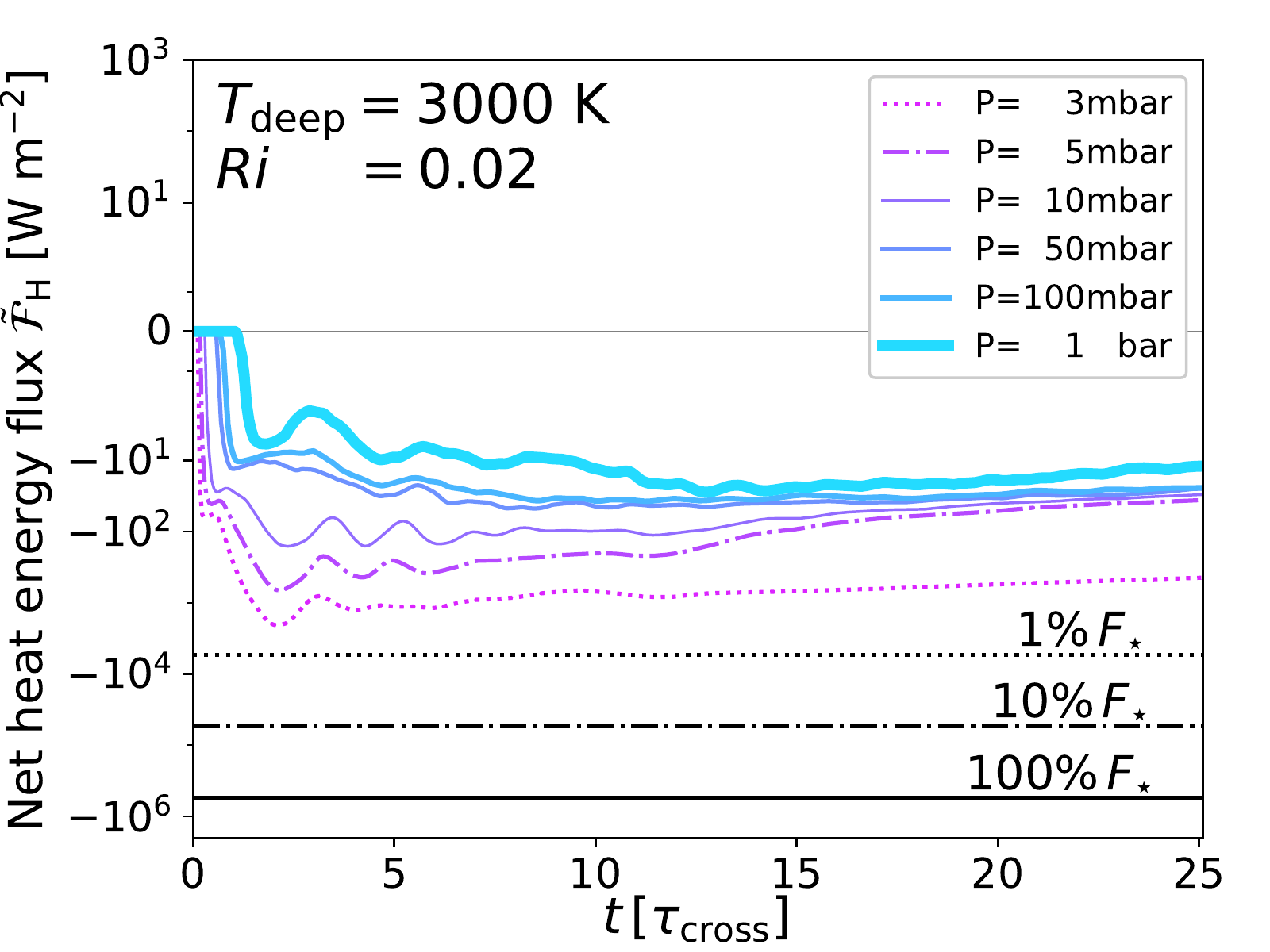}\hspace{-0.08in}
	\includegraphics[width=5.95cm]{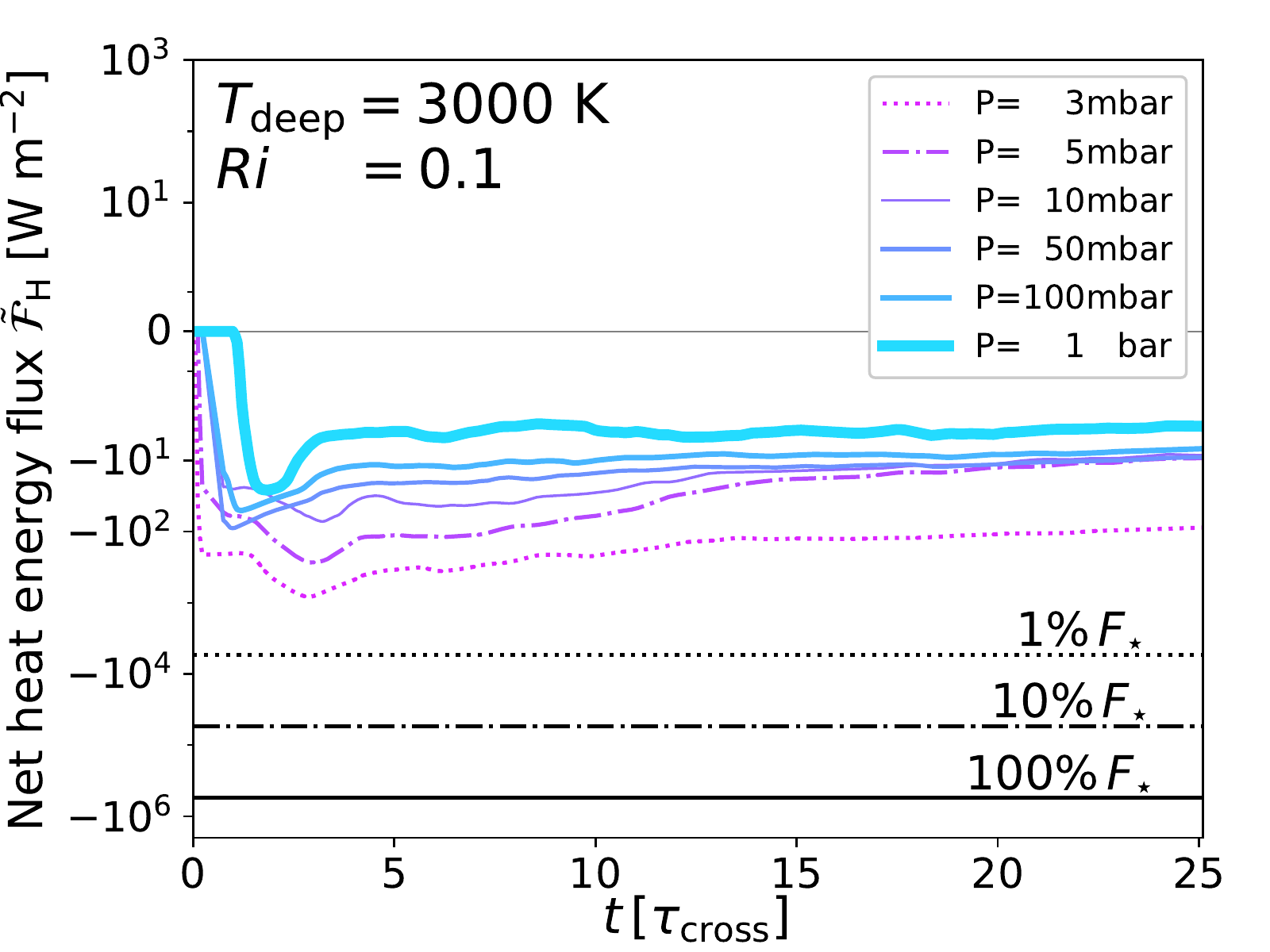}\hspace{-0.08in}
	\includegraphics[width=5.95cm]{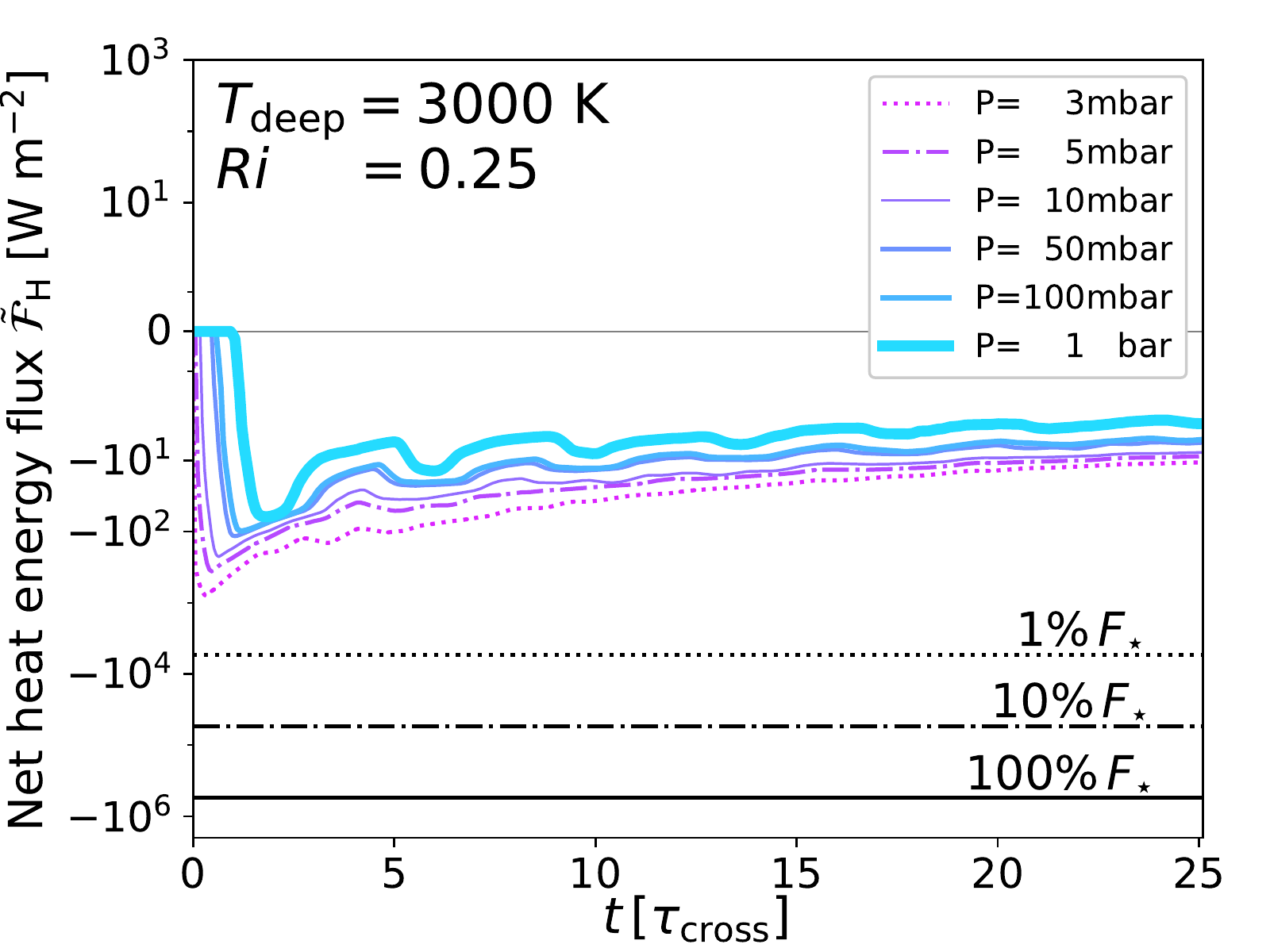}\hspace{-0.08in}
	\caption{Time evolution of the turbulence heat flux $\mathcal{\tilde{F}}_{\rm H}$ calculated using Equation \ref{eq:heatflux}  for $Ri=0.02$ (\textit{left} panel), $Ri=0.1$ (\textit{middle} panel) and $Ri=0.25$ (\textit{right} panel).}
	\label{fig:time_heatflux}
\end{figure*}

\begin{figure}
	\centering
	\includegraphics[width=8.2cm]{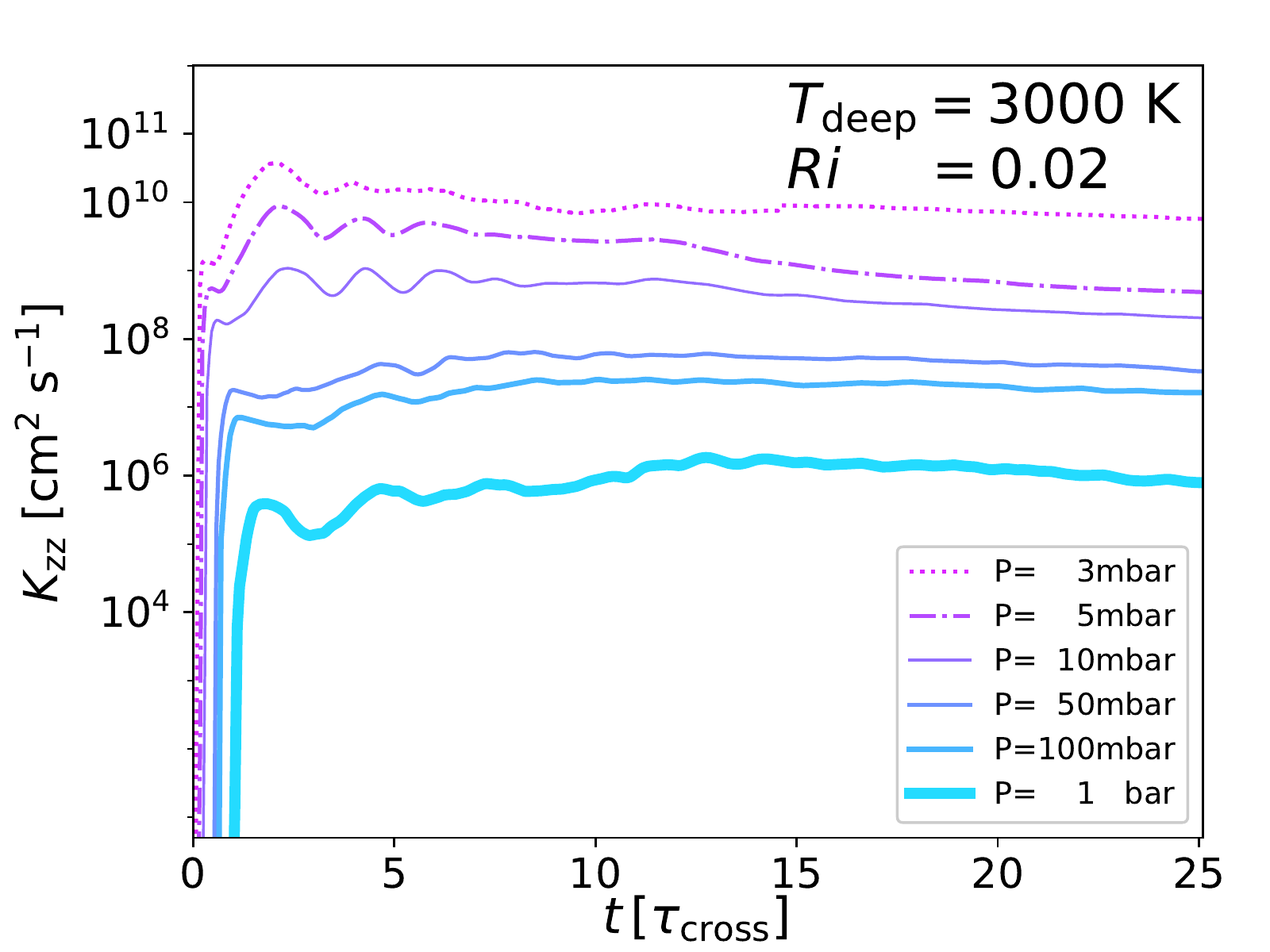}
	\includegraphics[width=8.2cm]{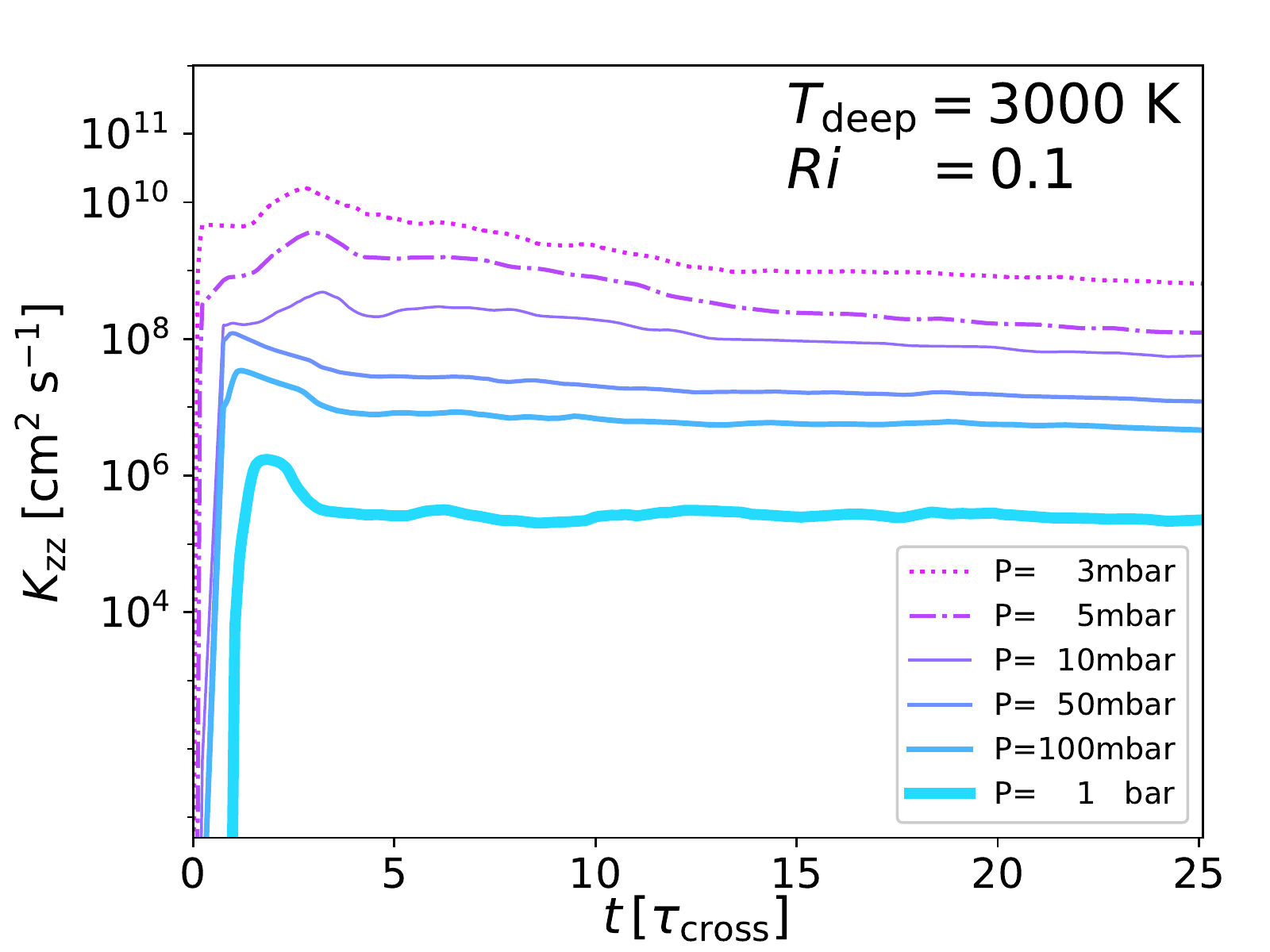}
	\includegraphics[width=8.2cm]{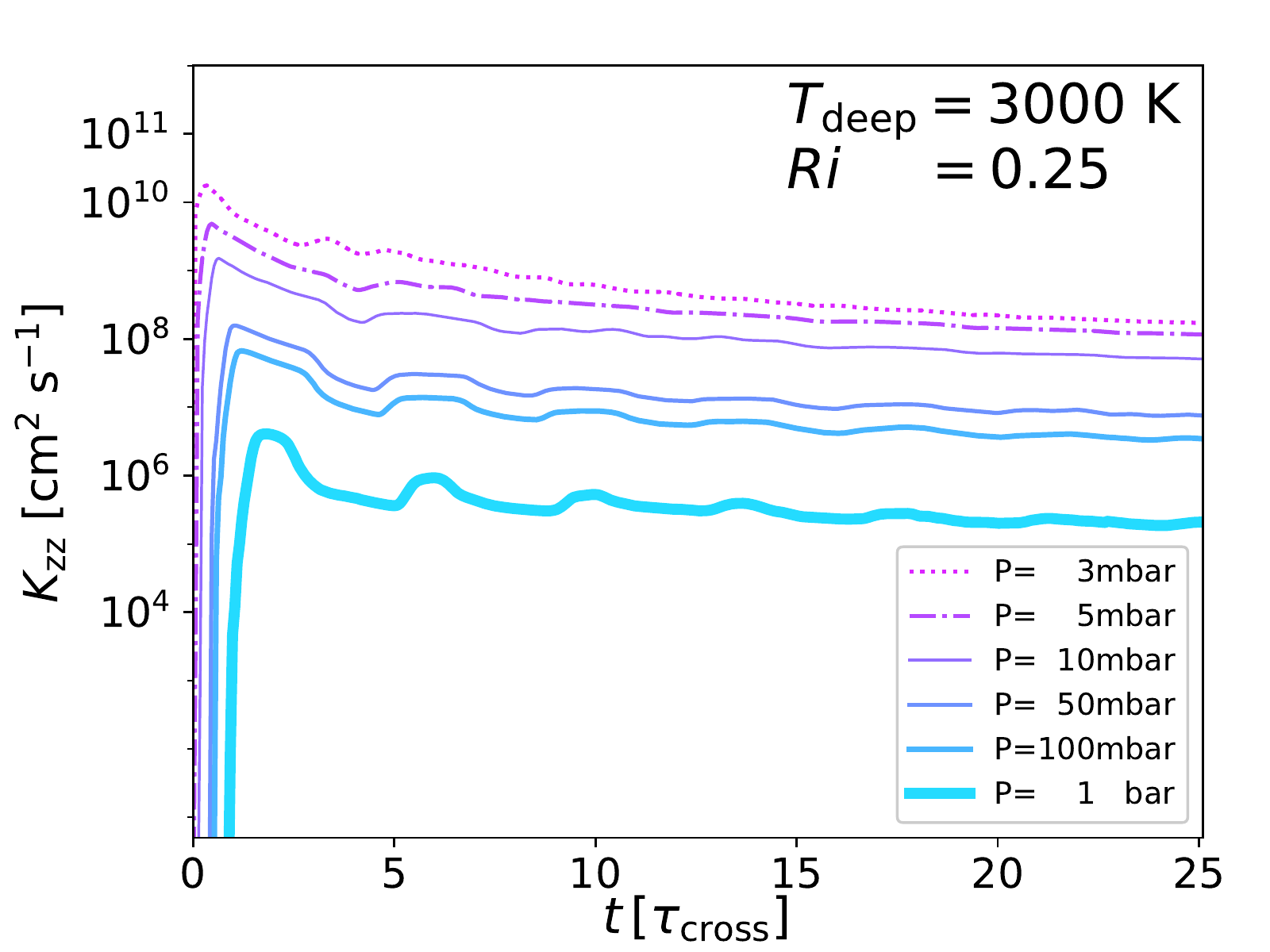}
	\caption{The time evolution of the eddy diffusion coefficient $K_{\rm zz}$ for $Ri=0.02$ (\textit{top} panel), $Ri=0.1$ (\textit{middle} panel) and $Ri=0.25$ (\textit{bottom} panel).}
	\label{fig:time_Kzz}
\end{figure}

\subsection{Kinetic energy and heat flux transport due to turbulence}
\label{sec:turbulenceflux}

Figures \ref{fig:time_turbkin} and \ref{fig:time_heatflux} show the time-averaged turbulence vertical kinetic energy flux $\tilde{\mathcal{F}}_{\rm KE}$ and the turbulence vertical heat flux $\tilde{\mathcal{F}}_{\rm H}$, respectively. In the flux figures, the three horizontal lines near the bottom indicate 1\% ($\simeq-10^{6}~\rm{W}~\rm{m}^{-2}$, solid horizontal line), 10\% ($\simeq-10^{5}~\rm{W}~\rm{m}^{-2}$, dot-dashed line) and 100\% ($\simeq-10^{4}~\rm{W}~\rm{m}^{-2}$, dotted line) of the incoming stellar flux $F_{\star}$ at $T=3000\K$. We first define an instantaneous kinetic energy flux $\tilde{F}_{\rm KE}$ \citep{Hannoun+1988}, while $\tilde{F}_{\rm H}$ \footnote{ \citealt{Hurlburt+1984} defines the heat flux such that positive $\tilde{F}_{\rm H}$ is directed downward. Notice the negative sign in their definition.} defines the heat flux as $\overline{ \rho c_{\rm p} v_{\rm z}T'}$ at $t$ and $P$ as follows,

\begin{align}
\tilde{F}_{\rm KE}(t,~P)&=\overline{\rho}\overline{\tilde{E}_{\rm KE}v^{'}_{\rm z}}, \label{eq:KineticEflux}\\
\tilde{F}_{\rm H}(t,~P)&=\overline{\rho  c_{\rm p} v_{\rm z}T'},
\label{eq:heatflux}
\end{align}
where $c_{\rm p}$ is the specific heat capacity at constant pressure, $c_{\rm P}=R/\nabla_{\rm ad}$. $\tilde{E}_{\rm KE}$ is the turbulence specific kinetic energy,

\begin{align}
\tilde{E}_{\rm KE}&=\frac{1}{2}({v'}_{\rm y}^{2}+{v'}_{\rm x}^{2}+{v'}_{\rm z}^{2}).
\label{eq:turbkinE}
\end{align}

Using then the instantaneous fluxes extracted from the output data at specific time intervals ($\Delta t/\tau_{\rm cross} \simeq 0.06$)\footnote{To maximize the code speed, the current default set-up of the code allows to print out the main state variables (e.g. $\rho$, $T$ and etc.) and some derived variables (e.g., entropy and etc.). We post-process the data for more model-specific variables, such as $\mathcal{\tilde{F}}$.}, we calculate a time average of the fluxes at a given $t$ and $P$ as, 
\begin{align}
\mathcal{\tilde{F}}(t,P)&=\frac{1}{t}\sum_{t'=0}^{t'=t}\tilde{F} (t',P)\Delta t.
\label{eq:netflux}
\end{align}

The turbulent kinetic energy flux $\mathcal{\tilde{F}}_{\rm KE}$ in Figure \ref{fig:time_turbkin} at $Ri=0.25$ (\textit{right} panel) is nearly zero throughout the atmosphere below the shear layer. On the other hand, for the lower $Ri$ (\textit{left} and \textit{middle} panels), the magnitudes of the fluxes $|\mathcal{\tilde{F}}_{\rm KE}|$ are smaller at higher $P$\footnote{We find that $\mathcal{\tilde{F}}_{\rm KE}$ in some deeper regions ($5\leq P\leq 10\Pmbar$) becomes positive. We believe that this is mostly due to small random fluctuation of $\tilde{F}_{\rm KE}$ around zero.}. The fluxes  at $P\simeq3\Pmbar$ for both  $Ri$'s are the largest and remain constant, but they are at most $0.01-0.001\%$ of $F_{\star}$. This means that the continuous shear forcing at the top keeps exciting eddy motions, but confined at $P\lesssim10\Pmbar$, then the turbulence kinetic energy rapidly dissipates into heat. Next, we quantify the turbulence heat energy flux $\mathcal{\tilde{F}}_{\rm H}$. 

Unlike the kinetic energy flux, the turbulence heat flux $\mathcal{\tilde{F}}_{\rm H}$ in Figure \ref{fig:time_heatflux} propagates downwards at all pressures with its magnitude smaller for higher $Ri$. However, this flux is not significant, roughly $\lesssim0.01\%$ of the incoming flux. Furthermore, $\mathcal{\tilde{F}}_{\rm H}$ for $Ri=0.25$ at all pressures  gradually converges to zero. Only for $Ri=0.01$, $\mathcal{\tilde{F}}_{\rm H}$ at $P\simeq3\Pmbar$ maintains a 0.1\% level. The fact that $\mathcal{\tilde{F}}_{\rm H}$ remains constant at later times means an unvarying inflow rate of the instantaneous flux $\tilde{F}_{\rm H}$ over the unit time (consider Equation \ref{eq:netflux} with constant $\tilde{F}_{\rm H}$). 

Overall, our results suggest that turbulence due to the shear motion near the top cannot lead to deep penetration of the energy flux, which remains confined in a vertical spatial scale of $\sim2H$. The atmosphere below $P=10\Pmbar$ is barely affected by the turbulent motion at $P\simeq 1-3\Pmbar$.

\subsection{Eddy diffusion coefficient}

 Our numerical experiments have shown that heat energy transport via turbulence by a forced shear layer is not a large-scale effect, but is rather locally confined within a vertical range of $\sim2H$. However, it is still worth quantifying the eddy diffusion coefficient $K_{\rm zz}$ for negative $F_{\rm H}$ in the atmosphere.  We estimate $K_{\rm zz}$ by combining equation (20) in \citet{YoudinMitchell2010},
\begin{align}
F_{\rm H}=-K_{\rm zz}\rho T\frac{dS}{dz},
\label{eq:heatflux_Kzz}
\end{align}
with the time-averged heat flux $\mathcal{\tilde{F}}_{\rm H}$ (Equations \ref{eq:heatflux} and \ref{eq:netflux}). These two equations give the following expression for $K_{\rm zz}$,
\begin{align}
K_{\rm zz}=\left|-\frac{\sum_{t'=0}^{t'=t}\overline{\rho c_{\rm P} v_{\rm z}T'}(\Delta t/t)}{\overline{\rho T\left(\frac{\Delta S}{\Delta z}\right)}}\right|,
\end{align}
where $\frac{dS}{dz}$ is estimated as follows,
\begin{align}
\frac{\Delta S}{\Delta z}(z=h)=\frac{S(z=h+\Delta z)-S(z=h-\Delta z)}{2\Delta z}.
\end{align}

\begin{figure}
	\centering
	\includegraphics[width=8.5cm]{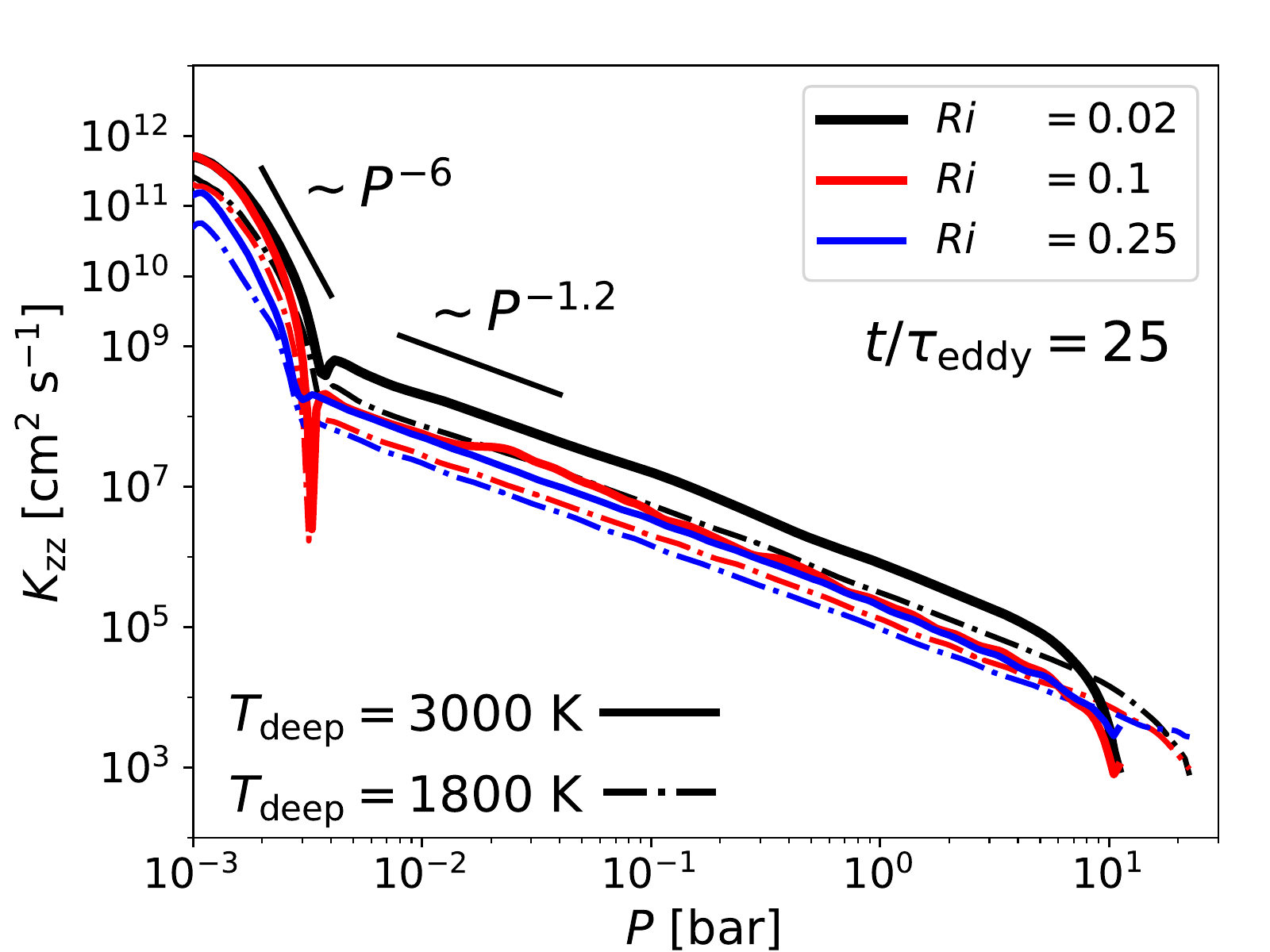}\hspace{-0.25in}
	\caption{$K_{\rm zz}$ as a function of $P$ for our atmosphere models with $T_{\rm deep}=3000\K$ (solid lines) and $T_{\rm deep}=1800\K$ (dot-dashed lines).}
	\label{fig:P_Kzz}
\end{figure}

 \begin{figure}
	\centering
	\includegraphics[width=8.4cm]{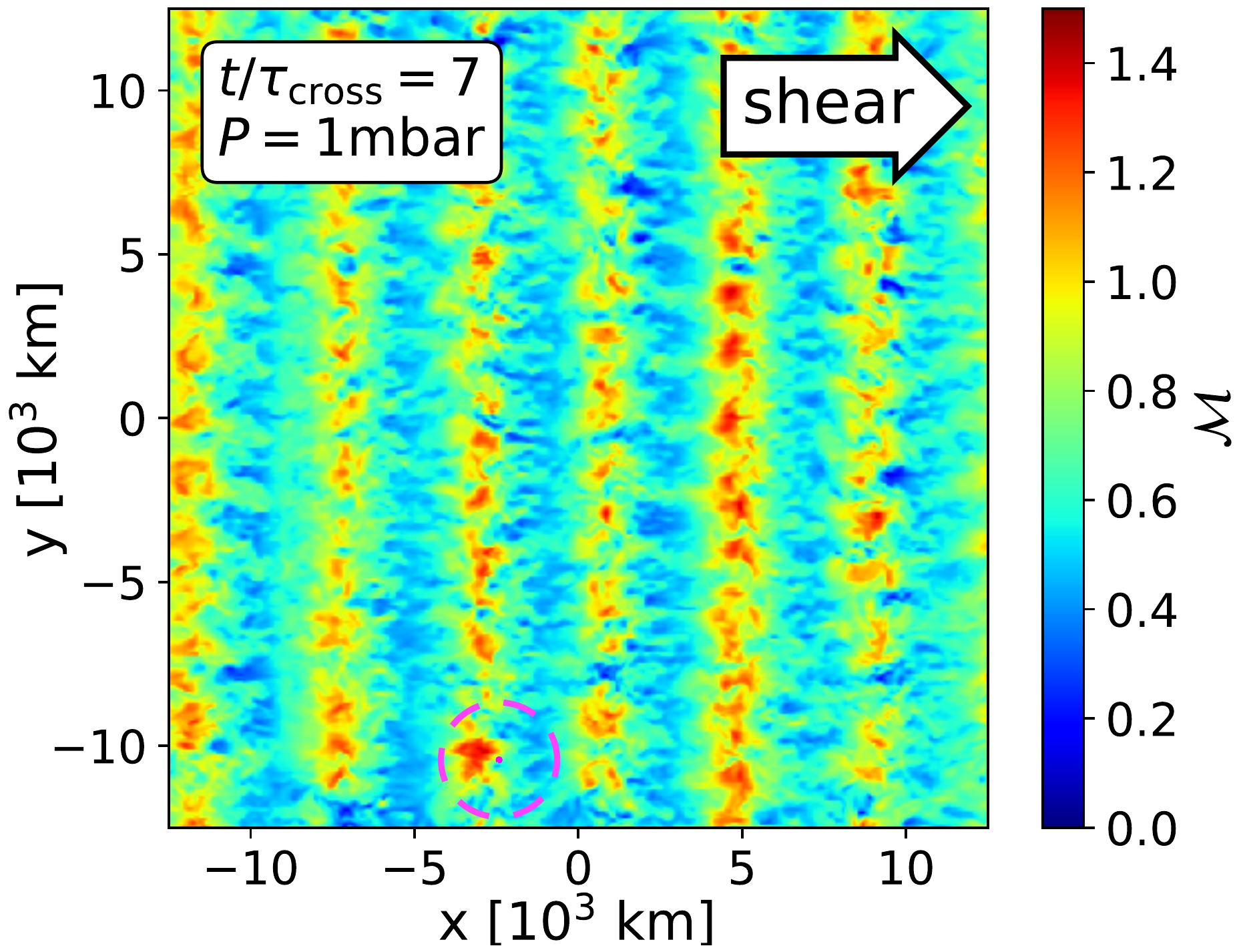}
	\includegraphics[width=8.4cm]{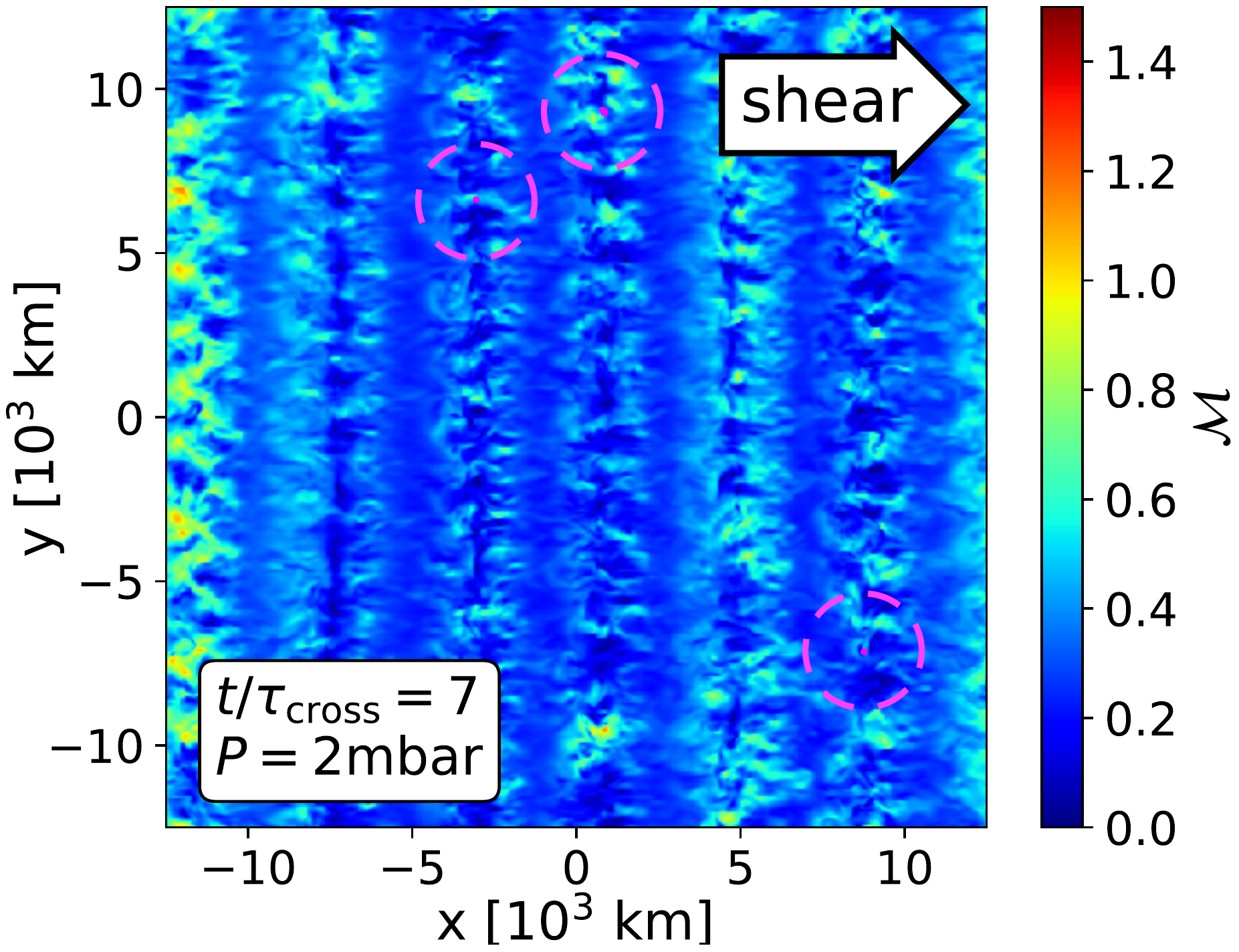}
	\caption{Horizontal slice plots for $\mathcal{M}$ in the atmosphere with $T_{\rm deep}=3000\K$ at $P=1\Pmbar$ (\textit{top} panel) and $P=2\Pmbar$ (\textit{bottom} panel) at $t=7~\tau_{\rm cross}$ when the initial $Ri=0.02$. The arrow at the top-right corner indicates the direction of continuous shear motion. The magenta circles indicate where shocks are detected. The plots are color-coded based on the magnitude of $\mathcal{M}$.}
	\label{fig:shockslice}
\end{figure}

Based on the time-averaged heat fluxes found in our models, we find (see Figure \ref{fig:time_Kzz}) that for $Ri=0.01$ (\textit{upper} panel), $K_{\rm zz}\simeq10^{8}-10^{10}\cm^{2}\s^{-1}$ at a few Pmbar pressures, decreasing down to $K_{\rm zz}\simeq10^{5}\cm^{2}\s^{-1}$ at $P=1\Pbar$. There is no significant difference between $Ri=0.1$ (\textit{middle} panel) and $Ri=0.25$ (\textit{bottom} panel), except for $P=3\Pmbar$. These values are reasonably consistent with \citet{Spiegel+2009}, where they estimate a $K_{\rm zz}\sim 10^{7}-10^{11}\cm^{2}\s^{-1}$ to be necessary to maintain sufficient TiO in the upper atmospheres ($P\simeq$ a few $\Pmbar$) for thermal inversion.

These estimates, however, do not clearly inform us on how the coefficient $K_{\rm zz}$ varies with pressure. To find the dependence of $K_{\rm zz}$ on $P$, we show $K_{\rm zz}$ as a function of $P$ for our atmospheric models with $T_{\rm deep}=3000\K$ (solid lines) and $T_{\rm deep}=1800\K$ (dot-dashed line) in Figure \ref{fig:P_Kzz}. We see that we have different $P$ dependences at $P\gtrsim 3-4\Pmbar$ and $P\lesssim 3-4\Pmbar$ for all the models. $K_{\rm zz}$  at  $P\lesssim 3-4\Pmbar$ dramatically declines, followed by a relatively mild decrease in the deeper atmosphere. Such steepness near the top is clearly due to chaotic eddy motions. Interestingly, $K_{\rm zz}$ for $Ri=0.02$ (black lines) and $Ri=0.25$ (blue lines) throughout the atmosphere have almost the same dependence on $P$, only differing by a factor of $\sim5$ in magnitude. On the other hand, $K_{\rm zz}$ for $Ri=0.1$ (red lines) shows a transitional behavior between that for $Ri=0.02$ (black lines) and $Ri=0.25$ (blue lines): the lines for $Ri=0.1$ are very close to those for $Ri=0.02$ at $P\lesssim 3-4\Pmbar$ whereas they are still lingering near the lines for $Ri=0.02$ at $P\gtrsim 3-4\Pmbar$. From this, we may be able to conjecture the following: 1) $Ri=0.1$ could be a characteristic value below which the heat flux starts effectively penetrating into the deeper region and 2) inflow of a heat flux in the inner region may occur via episodic jumps, rather than by a gradual growth. However, it is important to emphasize that our conjectures are made only based on our models, which cover a subset of the whole parameter space. In order to find more general trends of $K_{\rm zz}$ (e.g. how $K_{\rm zz}$ would increase in atmospheres for $Ri<0.01$, in particular whether it would gradually increase or whether there would be another lingering phase like the one we find for $Ri=0.25-0.02$), we need to explore a larger parameter space with different initial conditions, which we will leave for future work.

Assuming $K_{\rm zz}$ follows a power law of $P$ such that $K_{\rm zz}\propto P^{-\alpha}$, the following provides a fit for $K_{\rm zz}$, 
\begin{equation}
K_{\rm zz}\simeq\begin{cases}
5\times10^{8}\beta \left(\frac{P}{P_{\llcorner}}\right)^{-6}~~~\cm^{2}\s^{-1}   \hspace{0.5in} P<P_{\llcorner},\\
5\times10^{8}\beta \left(\frac{P}{P_{\llcorner}}\right)^{-1.2}~\cm^{2}\s^{-1}   \hspace{0.5in} P\geq P_{\llcorner},
\end{cases}
\label{eq:K_zz_fit}
\end{equation}	
where $\beta$ is a normalization factor, possibly depending on $T_{\rm deep}$ and the velocity gradient due to shear ($Ri$). As mentioned above, at least for $Ri=0.02$ and $Ri=0.25$, $\beta$ is a constant, differing by around 5. $P_{\llcorner}$ can be approximately found to be $P_{\llcorner}\simeq P$ at $z=z_{\rm 1\Pmbar}-1.5H$. 

The convergence of $K_{\rm zz}$ at the bottom ($P\simeq 10-30\Pbar$), along with the sharp decrease, in all our models is probably due to the boundary condition. This drop has been found in \citet{YoudinMitchell2010} for negative $\alpha$ (see their Figure 7), but near quite large pressures at which $\nabla=\nabla_{\rm ad}/2$. In our models, the pressure which satisfies the condition corresponds to $P\simeq 270\Pbar$ ($45\Pbar$) for $T_{\rm deep}=3000\K$ ($1800\K$), which is much higher than $P$ at the bottom.

\subsection{Shock formation}

Here we consider the formation of shocks in our atmosphere. We use a basic multi-dimensional shock detection algorithm \citep{ColellaWoodward1984,Colella1990} embedded in the code CASTRO to trace shocks. Overall, we find that shocks form, but they are sporadic (in space) and transient (in time). 

Shocks form within a range of $P\lesssim 2.5\Pmbar$ and the fraction of the areas where shocks are detected, defined as the ratio of the number of cells identified with shocks to the total number of cells at a given pressure, remains below the $10^{-3}-10^{-4}~\%$ level even in the most unstable atmosphere. We show horizontal slice plots for $\mathcal{M}$ in the atmosphere with $T_{\rm deep}=3000\K$ in Figure \ref{fig:shockslice} at $P=1\Pmbar$ (\textit{top} panel) and $P=2\Pmbar$ (\textit{bottom} panel) at $t=7~\tau_{\rm cross}$. This figure is made for the model with $Ri=0.02$, but the other models look very similar. The color indicates the magnitude of $\mathcal{M}$ as given in the color bar.
We also mark where shocks form using magenta dotted circles. They are local and scattered. 

\begin{figure}
	\centering
	\includegraphics[width=8.5cm]{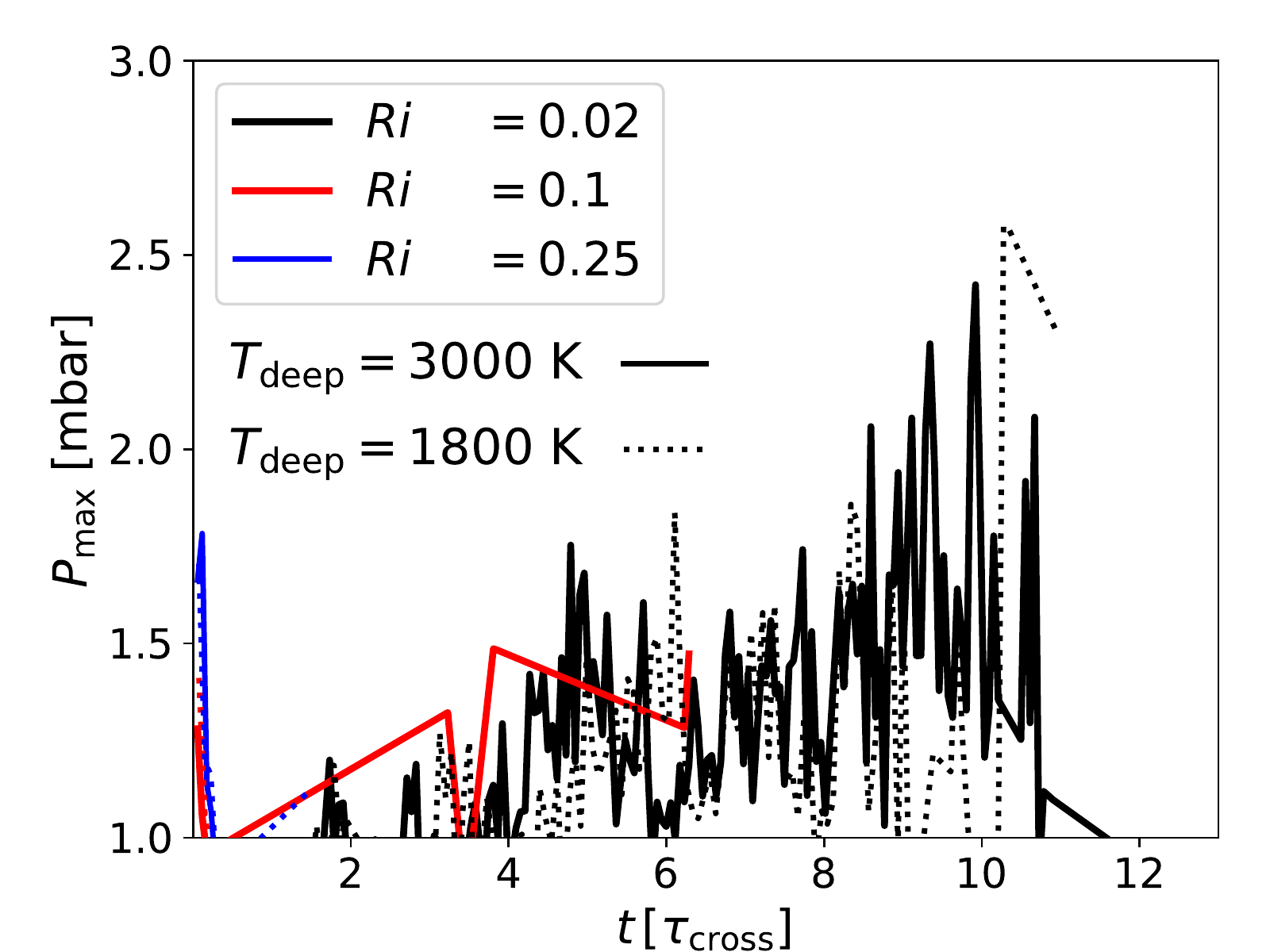}
	\caption{The maximum pressure at which shocks are detected ($P_{\rm max}$) as a function of time in our models with $T_{\rm deep}=3000\K$ (solid lines) and $T_{\rm deep}=1800\K$ (dotted line). The times at which the lines hit the bottom (i.e. at $P=1\Pmbar$) correspond to times when there are no shocks in the atmosphere. }
	\label{fig:shock_t}
\end{figure}

The shocks last longer in the atmosphere starting with smaller $Ri$, but they are no longer detected at $t\gtrsim 12~\tau_{\rm cross}$ for $Ri=0.02$. This is visualized in Figure \ref{fig:shock_t}. This figure shows the largest pressure at which shocks form ($P_{\rm max}$) as a function of time in our models with $T_{\rm deep}=3000\K$ (solid lines) and $T_{\rm deep}=1800\K$ (dotted line). The times when the lines cross the $P=1\Pmbar$ level correspond to moments when there are no shocks in the atmospheres. As clearly shown in the figure, the shocks cannot penetrate deeper than $P\simeq2\Pmbar$ and disappear rather quickly.

From the above, we conclude that shock formation is insignificant; therefore, shocks are not expected to affect the evolution of the atmospheres. Interestingly, even though \citet{Fromang+2016} investigated shocks based on different atmosphere models and criteria for shock formation, both studies suggest similar conditions for shock formation. Shocks are not found in their low resolution simulations. However, at resolution high enough to resolve finer structures of jets, they find instabilities which cause velocity fluctuations, ultimately transforming into weak shocks at $P\simeq$ a few $\Pmbar$. This trend (i.e. resolution dependence and weak shocks confined to lower pressures) is in good agreement with the findings from our simulations.

\begin{figure}
	\centering
	\includegraphics[width=8.3cm]{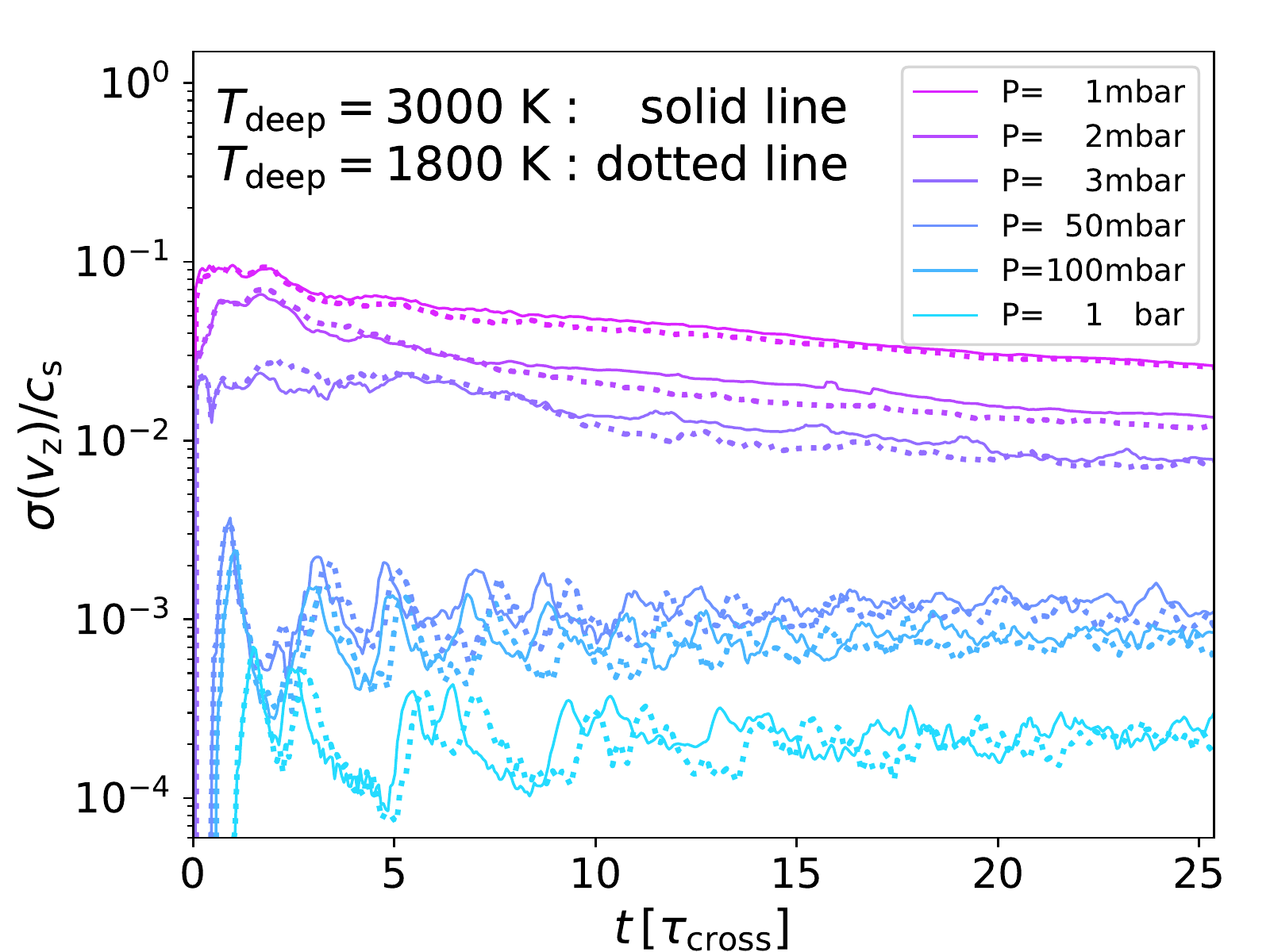}
	\includegraphics[width=8.3cm]{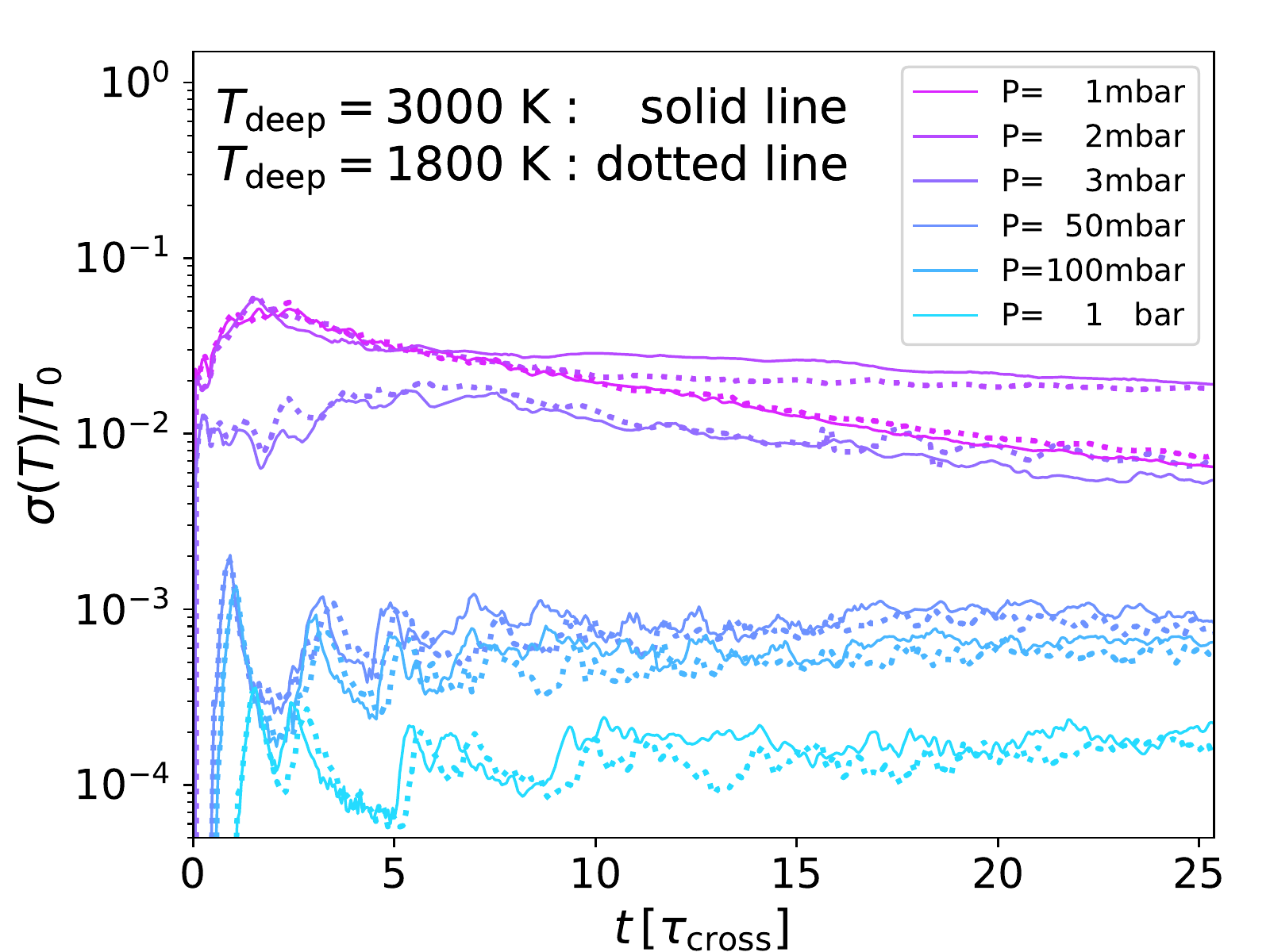}
	\includegraphics[width=8.3cm]{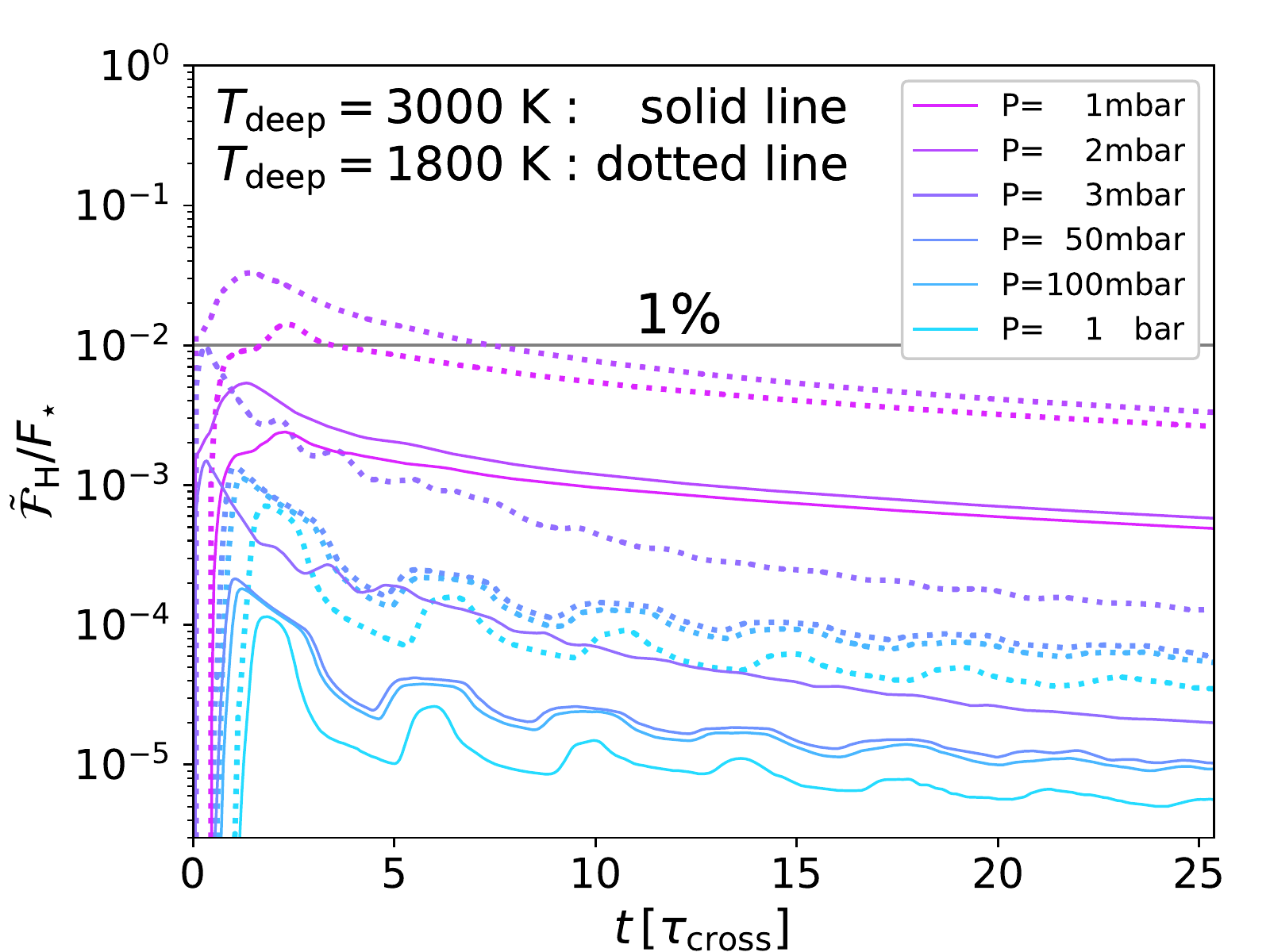}
	\caption{Root-mean-square ($\sigma$) of $v_{\rm z}$ and $T$, as well as the heat flux, normalized by $c_{\rm s}$, $T_{0}$, and $F_{\star}$ as a function of time for the two models with different $T_{\rm deep}$ at different pressures. The black horizontal line in the \textit{bottom} panel corresponds to 1\% of $F_{\star}$. }
	\label{fig:comparison_T}
\end{figure}

\begin{figure*}
	\hspace{-1.9in}
	\vbox to220mm{	\vspace{-0.7in} 
		\includegraphics[width=21.3cm]{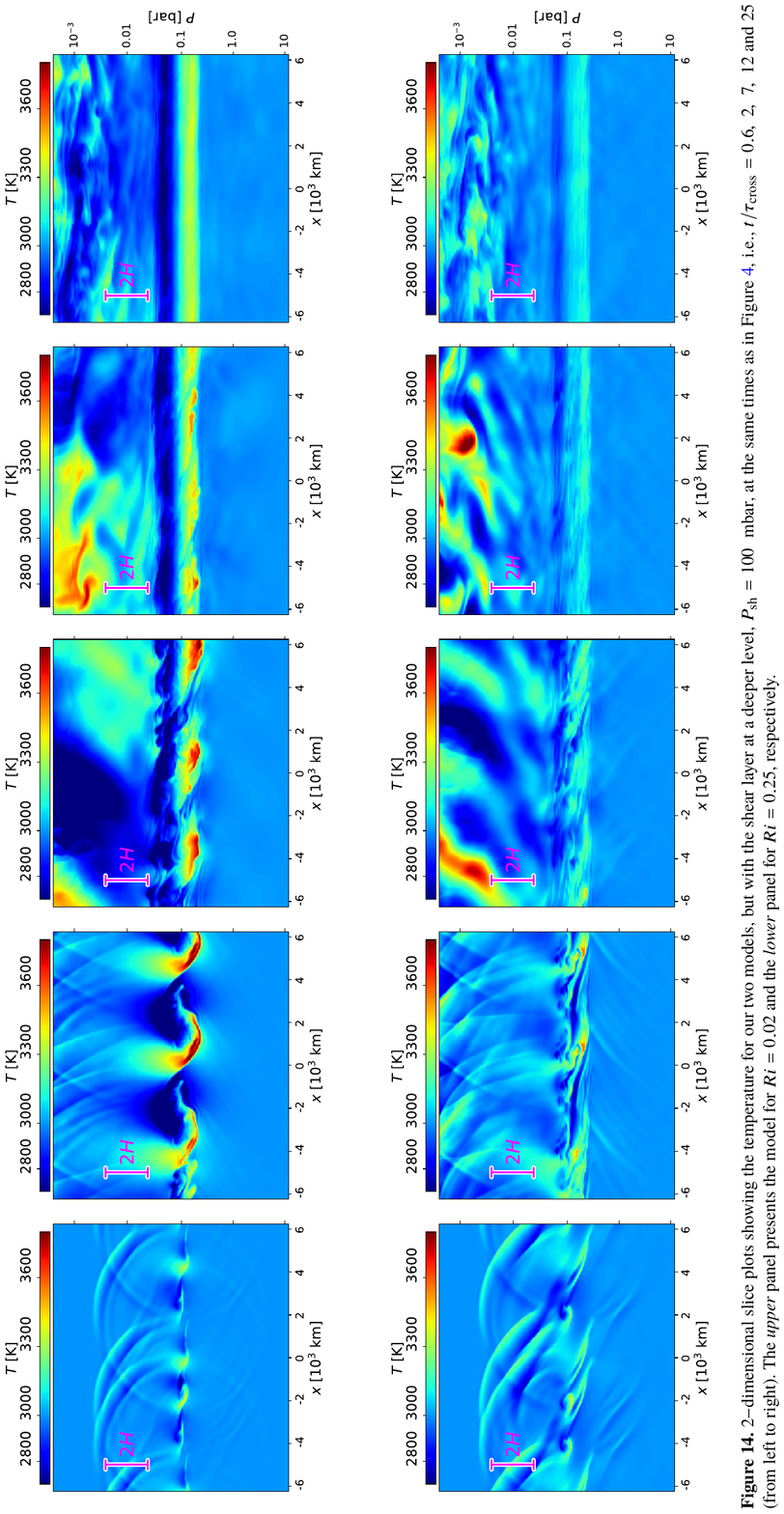}
		\caption{}
	}
	\label{fig:slice100mbar}
\end{figure*}

\begin{figure*}
	\centering
	\includegraphics[width=8.4cm]{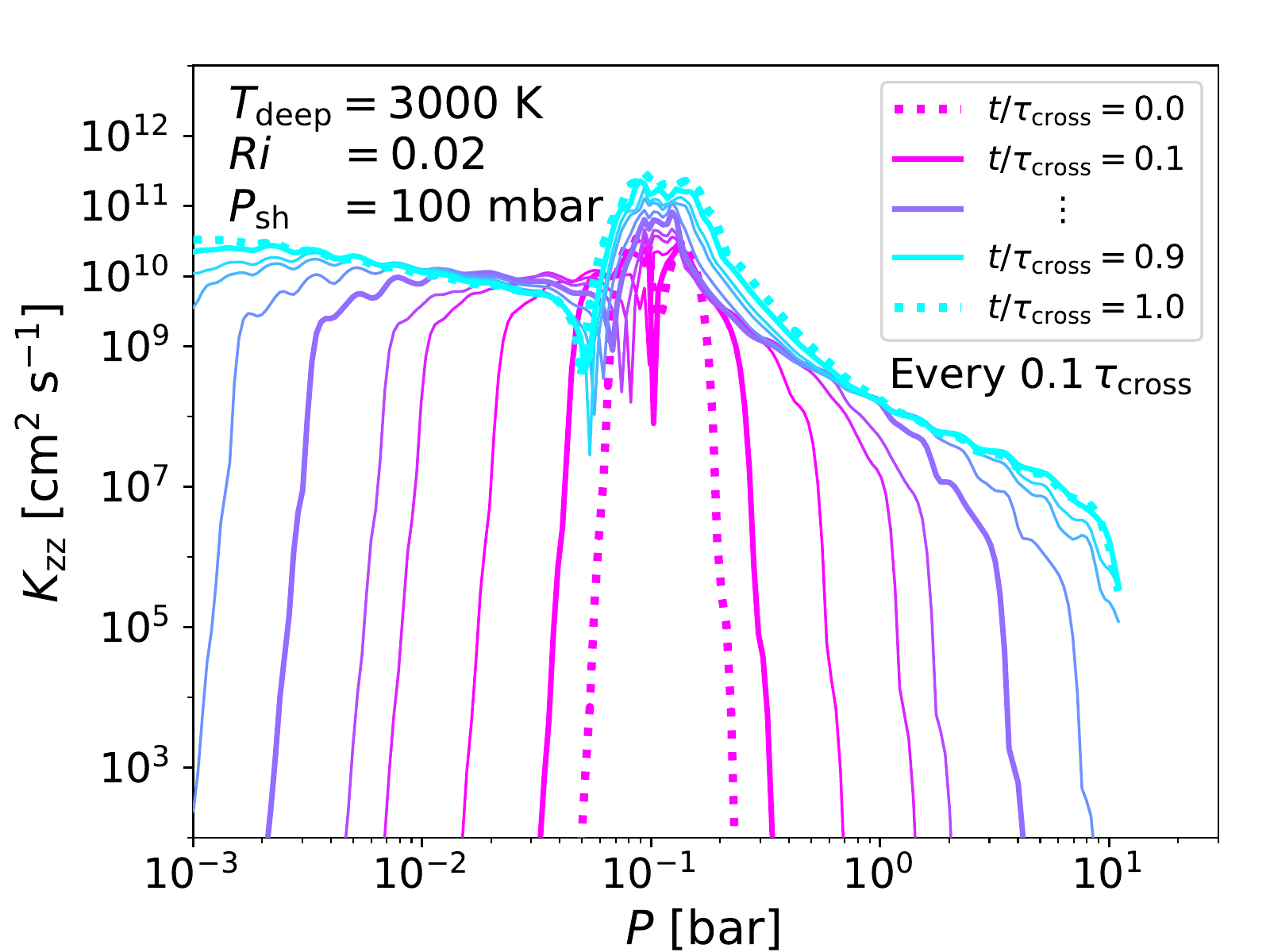}	
	\includegraphics[width=8.5cm]{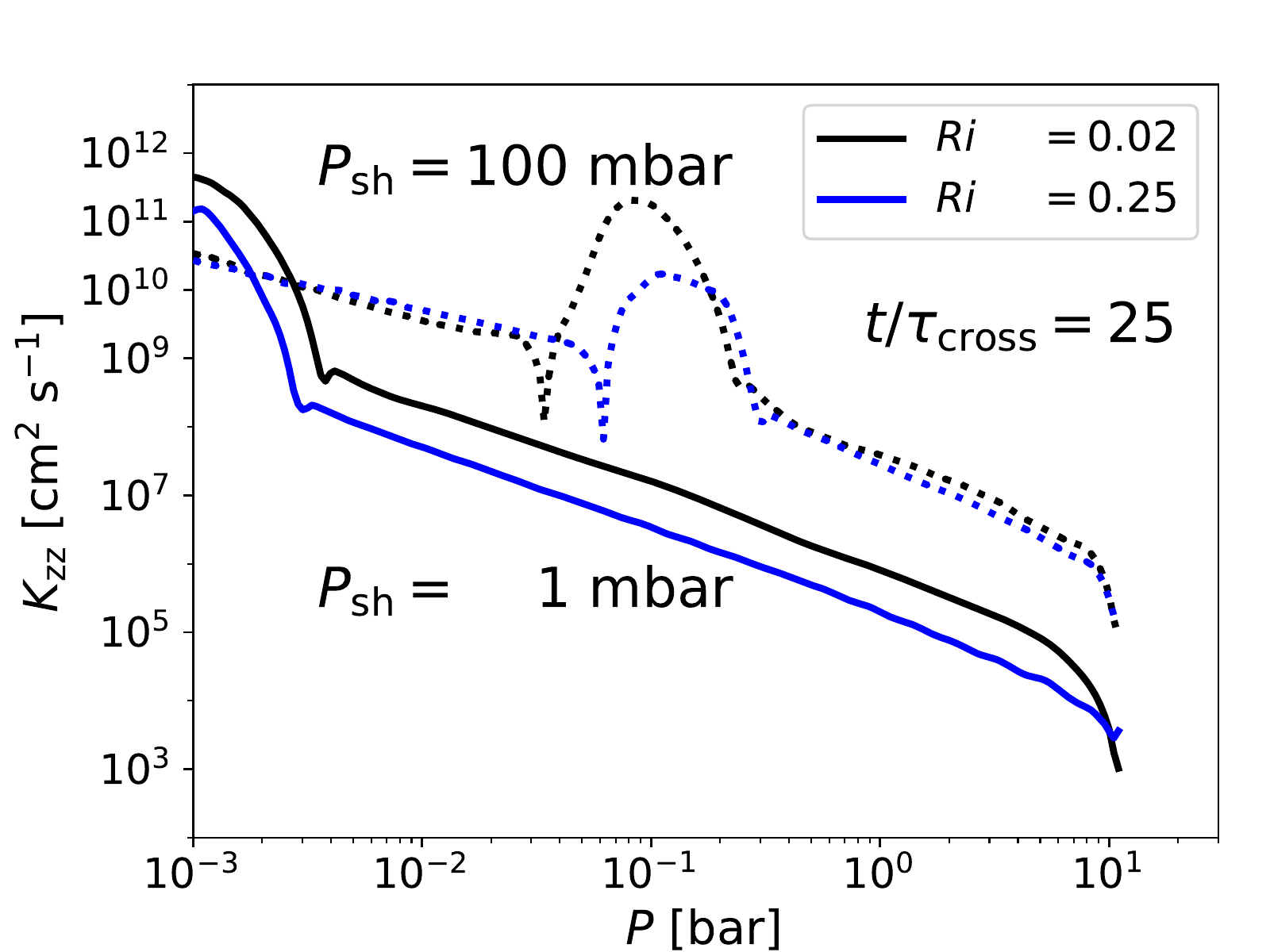}	
	\caption{\textit{Left} panel: time evolution of $K_{\rm zz}$ every $0.1~t/\tau_{\rm cross}$ for $0\leq t/\tau_{\rm cross}\leq1$ (the dotted lines indicate $t/\tau_{\rm cross}=0$ and $1$ and the solid lines for the intermediate times). \textit{Right} panel: $K_{\rm zz}$ as a function of $P$ for the atmospheres with different $P_{\rm sh}$ for $Ri=0.02$ (black lines) and $Ri=0.25$ (blue lines). The dotted lines indicate the models with $P_{\rm sh}=100\Pmbar$ and the solid lines those with $P_{\rm sh}=1\Pmbar$. }
	\label{fig:K_zz_100mbar}
\end{figure*}
\section{Discussion}
\label{sec:discussion}

\subsection{Comparison: different resolutions and the role of $T_{\rm deep}$}
\label{sec:T_comparison}

In this section, we compare our simulations between different resolutions and discuss the role of $T_{\rm deep}$.

\begin{enumerate}
	\item \textit{Resolution} 

As mentioned in \S \ref{sec:initialcondition}, to test the numerical reliability of our results, we performed every set of our simulations with two resolutions ($N_{\rm z}=1024$ and $512$).
	We find converging results between the high and low-resolution simulations, which adds more robustness to our results.  More specifically: in the low resolution simulations it is found that fluctuations in all the variables are generally larger at the initial times and inside the shear layer; however, when the atmosphere becomes steady, the final values of the variables are in a quite good agreement with the higher resolution simulations. One difference worth noting is the maximum depth at which shocks form, or $P_{\rm max}$. In the low resolution simulations, the depth extends to $P\simeq5\Pmbar$ and the shocks last longer, roughly by a factor of 2 than the higher resolution runs, and more shocks are detected (the fraction of atmospheric area containing shocks goes up to $10^{-1}\%$). Hence the formation of shocks can be overestimated if the resolution is not sufficiently small.

	\item \textit{$T_{\rm deep}$} 
	
The evolution of all the relevant variables mentioned so far, namely $T$, $v_{\rm z}$ and the energy fluxes, are almost identical\footnote{ Note that $\Delta l$ (with respect to $H$) for the models with different $T_{\rm deep}$ and their initial $T-P$ profiles (such as $P(z=0)$) are not exactly the same. The number of cells per scale height for $N_{\rm z}=1024$ (high resolution) is $21.8$ for $T_{\rm deep}=3000\K$
and $20.4$ for  $T_{\rm deep}=1800\K$.}. Among those variables, we show the RMS $\sigma$ of $v_{\rm z}$ and $T$ and the heat fluxes for the two models in Figure \ref{fig:comparison_T}. For a self-similar comparison, we normalize each value by a characteristic variable in the same dimension, such as $T_{0}$, $c_{\rm s}$ and $F_{\star}$. The black horizontal line in the \textit{bottom} panel indicates 1\% of $F_{\star}$. $T$ and $v_{\rm z}$ are found to be very comparable whereas $\mathcal{\tilde{F}}_{\rm H}/F_{\star}$ shows some discrepancies.  Indeed, it is because the magnitudes of $\mathcal{\tilde{F}}_{\rm H}$, not $\mathcal{\tilde{F}}_{\rm H}/F_{\star}$, for the two models are very similar. This may mean that, as long as the Mach number of the  horizontal shear motion at top is the same, the amount of heat flux reaching a certain pressure level is independent of $T_{\rm deep}$. This argument will need to be further explored in future work.

\end{enumerate}

\subsection{Turbulence in the deep regions}

\label{sec:turbulenceindeep}

So far, we have focused on turbulence initially created at $P_{\rm sh}\simeq1\Pmbar$. However, we cannot rule out the possibility that turbulence is generated more deeply. In stable stratified atmospheres, turbulence can be caused by a breakdown of internal buoyancy waves, like in the atmospheres of the Earth, Mars and Venus \citep{Izakov2001,Izakov2002}. In many global model simulations for hot Jupiters, it has been found that transonic zonal winds extend vertically down to $P\simeq1\Pbar$ \citep[e.g][]{Showman+2009,RauscherMenou2010,RauscherMenou2012,Fromang+2016}. For example, \citet{Showman+2009} find from their global 3-dimensional numerical simulations peak zonal wind speeds of $3.5\km\s^{-1}$ at $P\simeq 10-100\Pmbar$ (corresponding to $\mathcal{M}>1$ assuming $T=1200\K$) and \citet{Fromang+2016} find $Ri$ at $P\simeq 1\Pbar$ can be as low as $0.1-0.25$. This means that atmospheres at those pressure levels may also be subject to shear instabilities. In addition, based on evolution calculations with the MESA code, \citet{KomacekYoudin2017} studied the impact of internal heating on the radius evolution of hot Jupiters  by systematically varying the depth and intensity of internal heating. They find that heating at $P\gtrsim10\Pbar$ is required to keep hot Jupiters inflated as large as their observed radii. This also supports the importance of turbulence at deeper atmospheric levels.

To explore the role of turbulence in the deeper regions, we additionally perform two simulations for the hotter atmosphere ($T_{\rm deep}=3000\K$) with a shear layer at $P_{\rm sh}\simeq100\Pmbar$, instead of $P_{\rm sh}\simeq 1\Pmbar$. In this experiment, we only consider $Ri=0.25$ and $Ri=0.02$. All other model parameters, except for $P_{\rm sh}$, are identical to our fiducial models with $P_{\rm sh}\simeq 1\Pmbar$, including the continuous momentum input at $P=1\Pmbar$. We present slice plots in Figure {\color{blue}14} for the two models with $Ri=0.02$ (\textit{upper} panel) and $Ri=0.25$ (\textit{lower} panel) at the same times as in Figure {\color{blue}4}, i.e.,  $t/\tau_{\rm cross}=0.6,~2,~7,~12$ and $25$. Figures {\color{blue}4} and {\color{blue}14}  share the same color-coding scheme.

Similarly to the models with $P_{\rm sh}=1\Pmbar$, the temperature of gas near the shear layer becomes hotter as the kinetic energy of the gas dissipates into heat energy. Then the heat energy spreads out towards regions with relatively low $T$ from the shear layer.  Finally, the atmosphere becomes steady. The propagation of the heat energy can be visualized from how $K_{\rm zz}$ at each pressure level evolves over time. This is shown in the \textit{left} panel of Figure \ref{fig:K_zz_100mbar} for $Ri=0.02$. This panel shows $K_{\rm zz}$ every $0.1~t/\tau_{\rm cross}$ for $0\leq t/\tau_{\rm cross}\leq1$ (the dotted lines indicate $t/\tau_{\rm cross}=0$ and $1$, while the solid lines mark the intermediate times). 

There are two points worth noting: 1) One outcome which has not been seen in the fiducial models, but it is seen in this experiment, is that the regions above the shear layer go through larger increases in $T$ (see high temperatures at $P<100\Pmbar$ in Figure {\color{blue}14} compared to those at $P>100\Pmbar$). This is because a relatively small amount of heat energy is necessary to increase the temperature in a less dense region. 2) Unlike our fiducial models with $P_{\rm sh}=1\Pmbar$, we find that the values of $K_{\rm zz}$ for $Ri=0.02$ and $Ri=0.25$ outside the shear layer are comparable when a deeper shear layer is considered. This can be explained from trade-offs between efficiency of heat energy conversion via turbulence and the total kinetic energy budget which can dissipate into heat energy: according to our shear prescription, the total initial momentum (kinetic energy) of the shear layer increases as $Ri$. Since we do not consider the continuous momentum input near the shear layer in these simulations, the total kinetic energy budget for $Ri=0.02$ which can dissipate to heat energy is (five times) smaller than that for $Ri=0.25$. Therefore, even though the heat energy can be converted

via turbulence more efficiently in a more unstable atmosphere with $Ri=0.02$, however it is limited by the smaller kinetic energy budget contained in the shear layer. On the other hand, for $Ri=0.25$, the conversion efficiency is lower, but the shear layer has a larger reservoir of kinetic energy.

Overall, a larger heat flux can reach deeper regions when eddy motions are created at larger pressures. In this additional experiment with $P_{\rm sh}=100\Pmbar$, $\mathcal{\tilde{F}}_{\rm H}$ at $P\simeq1\Pbar$ ($10\Pbar$) becomes comparable to $\sim0.1$\% (0.01\%) of $F_{\star}$ for both $Ri$'s; these values are larger than those with $P_{\rm sh}=1\Pmbar$ by roughly two orders of magnitude. As a result, as shown in the \textit{right} panel of Figure \ref{fig:K_zz_100mbar}, $K_{\rm zz}$ with higher $P_{\rm sh}$ (dotted lines) is larger than that with $P_{\rm sh}=1\Pmbar$ (solid lines) by several orders of magnitude throughout the atmosphere, except near $P_{\rm sh}\simeq1\Pmbar$. However, no significant difference in $\mathcal{\tilde{F}}_{\rm H}$ at $P\simeq1\Pbar$ is found between the two $Ri$'s. 
It is interesting to note that one can recover the dotted lines (higher pressure) by translating towards higher $P$ 
the lower-pressure (solid) lines by an amount comparable to the
ratio between the two $P_{\rm sh}$ values. Note that $K_{\rm zz}$ at $P_{\rm sh}\simeq1\Pmbar$ is somewhat larger, probably due to the continuous shear motion in the top layers. We will discuss this in more detail in the following section.

These additional results strengthen and broaden our argument that the effect of turbulence on the atmosphere below where eddies form is local, whereas it can cause a spatially large impact on the thermal evolution in the regions above it. Therefore, what is more important for effective heat  energy transfer into deeper regions via turbulence is probably where an atmosphere becomes unstable, rather than how unstable it is. Our results further add another aspect, which is that deep shear instabilities can significantly affect the atmosphere above where eddies are created.

\begin{figure*}
	\centering
	\includegraphics[width=8.4cm]{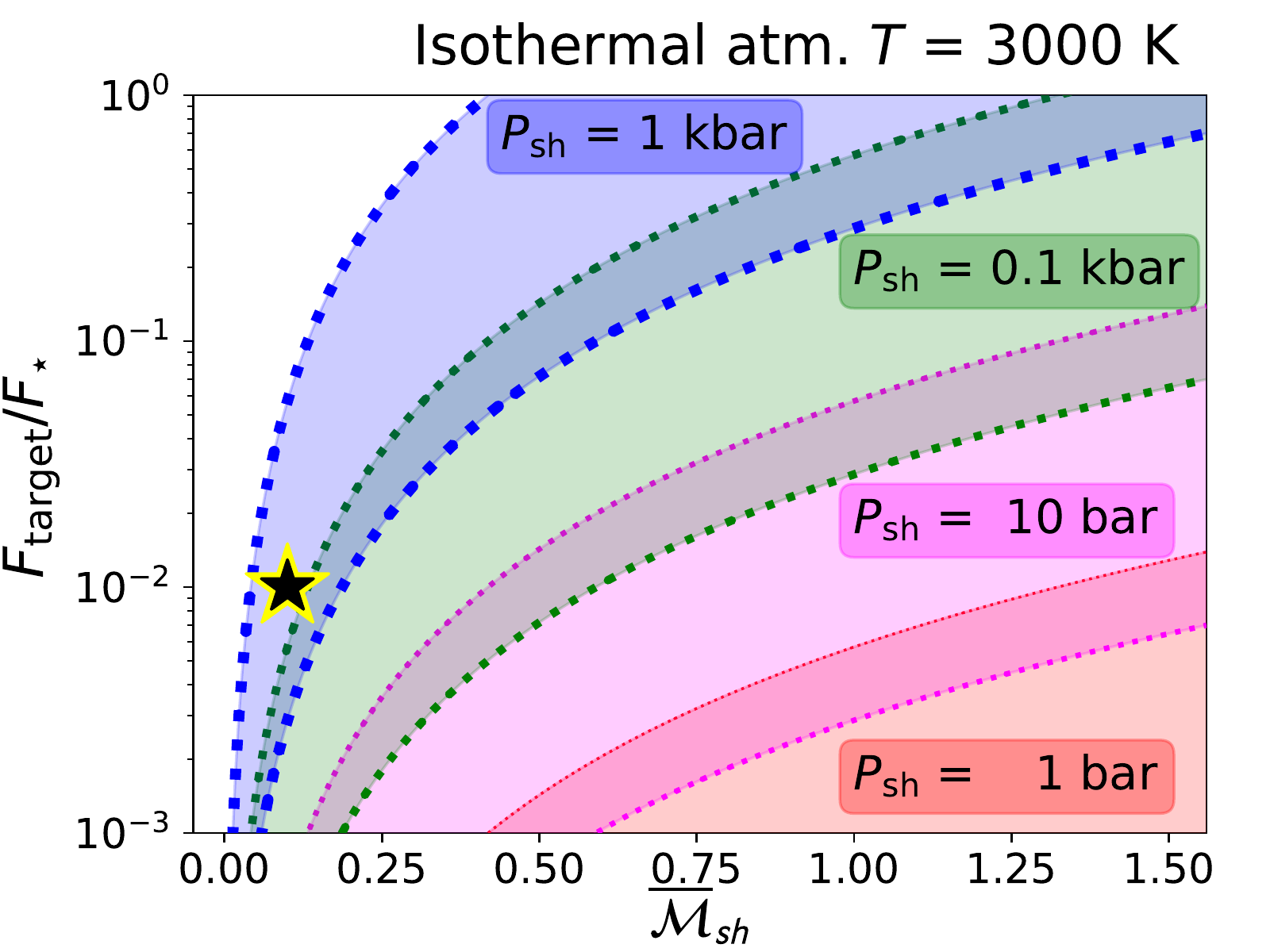}	
	\includegraphics[width=8.5cm]{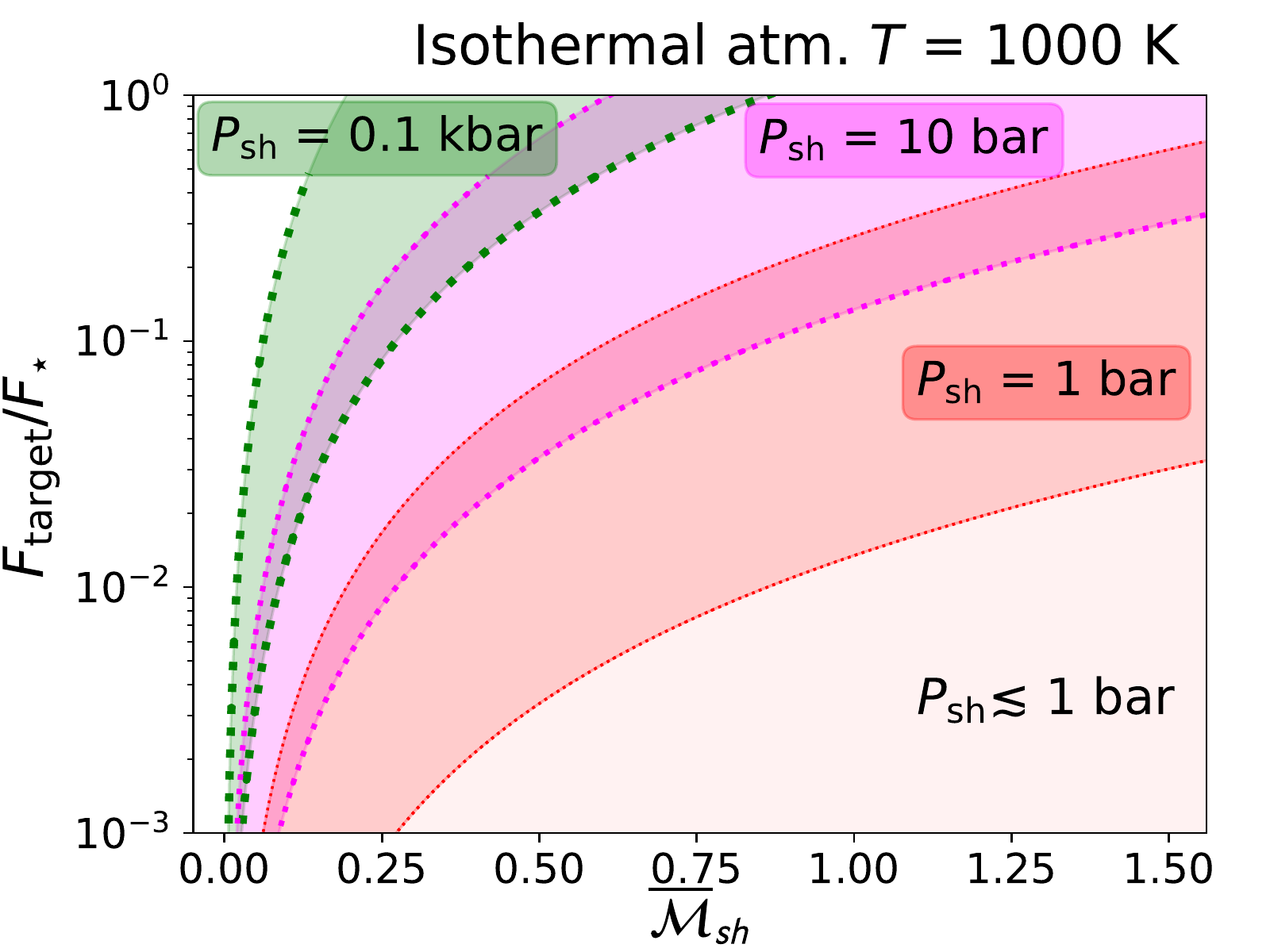}	
	\caption{Local shear-driven heat energy flux $F_{\rm target}$ with respect to the incoming stellar flux $F_{\star}$, in isothermal atmospheres with $T=3000\K$ (\textit{left} panel) and $T=1000\K$ (\textit{right} panel) as a function of jet stream velocity ($\overline{\mathcal{M}}_{\rm sh}$) at different pressure levels. The flux ratio $F_{\rm target}/F_*$ is estimated using Equation~(\ref{eq:analytic}), hence appearing as a region at a given $P$. The darker zones between regions indicate where two zones overlap. The dotted lines demarcating the boundary of each region correspond to the upper and lower limits of the flux ratio at that pressure level. The lines become thicker as $P$ increases ($\nwarrow$ direction). Note that, for the cases considered in our simulations, i.e. $P_{\rm sh}=1\Pmbar$ and $100\Pmbar$, the ratio $F_{\rm target}/F_*$ is smaller than $10^{-3}$. The star in the \textit{left} panel indicates a jet stream velocity which would generate a heat flux of 1\% $F_{\star}$ near the RCB.}
	\label{fig:Analytic}
\end{figure*}

\section{Analytic relations for $P_{\rm sh}$ and $K_{\rm zz}$}

Based on our results, i.e. the local effect of turbulence, we can now find some useful analytic relations for $K_{\rm zz}$ and identify a characteristic minimum pressure of a shear layer at which heat energy created by turbulence can be comparable to the energy necessary to account for the observed radii of hot Jupiters.

Let us consider a shear layer with a height $\Delta z_{\rm sh}$, from $z_{\rm b}$ at the bottom to $z_{\rm t}$ at the top of the layer, in an isothermal region of a planet. For simplicity, we assume a constant velocity gradient $dv/dz=\xi=\Delta v_{\rm sh}/\Delta z_{\rm sh}$, like the assumption made for our models. Then the shear velocity at $z$ is $v_{\rm sh}(z)=(\Delta v_{\rm sh}/\Delta z_{\rm sh}) (z-z_{\rm b})$. Accordingly, the mean kinetic energy density $\overline{E}_{\rm KE}$ in the shear layer can be estimated as follows,

\begin{align}
\overline{E}_{\rm KE}&=\frac{1}{2\Delta z_{\rm sh}}\int_{z_{\rm b}}^{z_{\rm t}}\rho(z') ~v_{\rm sh}^{2}(z')~dz'\nonumber\\
&=\frac{1}{2\Delta z_{\rm sh}}\rho(z=z_{\rm b})~\xi^{2}\int_{0}^{\Delta z_{\rm sh}}(z')^{2}e^{-\frac{z'}{H}} dz'\nonumber\\
&=\rho(z=z_{\rm b})~\xi^{2}H^{2}\left[\frac{H}{\Delta z_{\rm sh}}-e^{-\frac{\Delta z_{\rm sh}}{H}}\left(\frac{H}{\Delta z_{\rm sh}}+1+\frac{\Delta z_{\rm sh}}{2H}\right)\right],
\end{align}
where we have used an isothermal density profile for $\rho(z)$. 
We then assume that the kinetic energy in the shear layer is ultimately converted into heat energy via turbulence and is transported downwards at $v_{\rm z,sh}$, i.e. $\mathcal{\tilde{F}}_{\rm H}\simeq \overline{E}_{\rm KE}\,v_{\rm z,sh}$. For an energy flux $F_{\rm target}=f F_{\star}=f \sigma_{\rm SB} T^{4}$, with a constant $f\leq1$, necessary to keep the planet bloated, we find the following relation,

\begin{align}
\rho(z=z_{\rm b})~(\Delta v_{\rm sh})^{2} v_{\rm z,sh}&\simeq \frac{F_{\rm target}}{\left[\frac{H}{\Delta z_{\rm sh}}-e^{-\frac{\Delta z_{\rm sh}}{H}}\left(\frac{H}{\Delta z_{\rm sh}}+1+\frac{\Delta z_{\rm sh}}{2H}\right)\right]}\left(\frac{\Delta z_{\rm sh}}{H}\right)^{2}\nonumber\\
&\simeq\frac{\beta^{2}~F_{\rm target}}{\left[\beta^{-1}-e^{-\beta}(\beta^{-1}+1+0.5\beta)\right]},
\end{align}
where $\beta=\Delta z_{\rm sh}/{H}>0$ and $v_{\rm z,sh}$ indicates a typical vertical velocity of eddy motions. Strictly speaking, $\rho(z=z_{\rm b})$ is the density at the bottom of the shear layer  and $\Delta v_{\rm sh}$ is the velocity at the top of the layer (since we assume $v_{\rm sh}(z=z_{\rm b})=0$), but they can be loosely interpreted as the average location and velocity of the shear layer, simply denoted by $\overline{\rho}_{\rm sh}$ and $\overline{v}_{\rm sh}$, respectively.
Our simulations suggest that $\beta\simeq 1-2$ and $v_{\rm z}/c_{\rm s}\simeq\sigma(v_{\rm z})/c_{\rm s}\simeq 10^{-3}-10^{-4}$ outside the shear layer (see Figure \ref{fig:comparison_T}), leading to,
\begin{align}
\overline{\rho}_{\rm sh}~(\overline{v}_{\rm sh})^{2} = \overline{P}_{\rm sh}\left(\frac{\overline{v}_{\rm sh}}{c_{\rm s}}\right)^{2} \simeq (10^{4}-10^{5})~f~\sigma_{\rm SB}~T^{4}c_{\rm s}^{-1}.
\end{align}
We rewrite this as, 

\begin{align}
\overline{P}_{\rm sh} \simeq (2-35)\Pbar ~(\overline{\mathcal{M}}_{\rm sh})^{-2}\left(\frac{f}{0.01}\right)\left(\frac{T}{3000\K}\right)^{7/2}.
\label{eq:analytic}
\end{align}

The above equation implies that, once turbulence is created at $P\simeq {\rm a ~few}-35\Pbar$ by equatorial jets with $\mathcal{M}\simeq 1$, it can lead to a heat energy transport with an amount as much as $1\%$ of $F_{\star}$. But this is likely to be a local energy input. For example, in this particular case, since it is still far from the RCB on a level of $1$ kbar, a smaller amount of heat energy would be transferred to the RCB. Using this relation, one can find a typical or minimum jet velocity ($\overline{\mathcal{M}}_{\rm sh}$) at any given $P$ (e.g. $P\simeq P_{\rm RCB}$ or less) and $T$ to achieve $F_{\rm target}$ directly "near" the needed pressure level. The relation between $f=F_{\rm target}/F_{\star}$ and the jet velocity for different pressure levels above the RCB is visualized in Figure \ref{fig:Analytic}.  As an example, eddies due to jet streams with $\overline{\mathcal{M}}_{\rm sh}\simeq 0.1$ at $\overline{P}_{\rm sh}=P_{\rm RCB}\simeq 1$ kbar (marked with a '$\star$' in the left panel of Figure \ref{fig:Analytic}) would generate a heat energy flux of $1\%~F_{\star}$ at that pressure level.

Going one step further, we can find an analytic relation for $K_{\rm zz}$. Using Equation \ref{eq:heatflux} (or Equation 20 in \citealt{YoudinMitchell2010} with $\nabla\simeq 0$) with $F_{\rm H}\simeq \overline{\rho}_{\rm sh}\overline{v}_{\rm sh}^{2}\sigma(v_{\rm z,sh})$, we find that,

\begin{align}
K_{\rm zz}&\simeq \frac{\overline{\rho}_{\rm sh}\overline{v}_{\rm sh}^{2}\sigma(v_{\rm z,sh})}{\overline{\rho}_{\rm sh} g}\simeq\frac{(\overline{v}_{\rm sh})^{2}\sigma(v_{\rm z,sh})}{g}\nonumber\\
&\simeq 4~(\overline{v}_{\rm sh})^{2} \left(\frac{\sigma(v_{\rm z,sh})}{10^{-2}c_{\rm s}}\right)\left(\frac{1000\cm\s^{-2}}{g}\right)\nonumber\\
&\propto (\overline{v}_{\rm sh})^{2}.
\end{align}
From this, we can see that the important factor in determining $K_{\rm zz}$ at the shear layer is probably $(\overline{v}_{\rm sh})^{2}$, or the specific kinetic energy of eddy motions. Outside the shear layer, $K_{\rm zz}$ would extend following the power law of $P^{-1.2}$ (see Equation \ref{eq:K_zz_fit}). This also explains why we find similar $K_{\rm zz}$ at different $\overline{P}_{\rm sh}$, as we briefly mentioned in \S\ref{sec:turbulenceindeep}, since we always assume $\overline{v}_{\rm sh}\simeq c_{\rm s}$, $K_{\rm zz}\simeq (10^{11}-10^{12})\cm^{2}\s^{-1}$ at $T=3000\K$ at $\overline{P}_{\rm sh}$. However,  the location of the sheer layer  ($\overline{P}_{\rm sh}$) is also important for the amount of heat flux transported, since $\mathcal{\tilde{F}}_{\rm H}\propto \rho(\overline{P}_{\rm sh})~K_{\rm zz}$, i.e. there is a direct dependence on the (location-dependent) kinetic energy, rather than simply on the specific kinetic energy.

To summarize, we suggest that the key factors in determining $\mathcal{\tilde{F}}_{\rm H}$ and $K_{\rm zz}$ are the kinetic energy and the specific kinetic energy at the shear layer, respectively.

\section{Caveats}

Our suite of hydrodynamics simulations, modelling turbulence and shocks in the atmospheres of hot Jupiters, show that their effects on the transport of heat energy fluxes are local in space and transient in time. As we discussed in \S\ref{sec:T_comparison}, our results for the effects of turbulence and shocks are reasonably robust. However, in the following we point out two caviats which will require future investigation.

\begin{enumerate}
\item{\textit{Radiative effects}}

	It has been shown in  global circulation models that shear motions at different pressure levels are triggered by east-west stream motions as the day side gets irradiated more than the night side. Furthermore, as mentioned above, the location of the RCB is closely related to where the radiative heating and cooling are balanced. All of this implies that the evolution of the planet atmospheres is governed by complicated physics of radiation, cooling and hydrodynamics.

In our study, we mimic the shear forcing by considering a momentum source. This allows us to better focus on the hydrodynamical effects of turbulence but we may miss some of the impact of radiative transfer and cooling. 
For example, it is possible that radiative transfer may be smoothing out some of the temperature gradients near the optically thin regions seen in our simulations (e.g. Figures {\color{blue}4} and {\color{blue}14}). The radiative time scale near $P\simeq 1-100\Pmbar$ (e.g., $\sim10^{3}-10^{4}\s$ in Figure 4 of \citealt{Showman2002}) can be comparable to the 
eddy evolution timescale in our simulations. That means that in some cases the cooling could produce non-negligible effects on the dynamics of the atmosphere. This will be investigated
in future studies.

\item{\textit{Vertical location of RCB, $P_{\rm RCB}$}}

The internal entropy is an important parameter for the vertical location of the radiative-convective boundary, or $P_{\rm RCB}$. In general, $P_{\rm RCB}$ increases with the internal entropy. In this study, we only assume the same internal entropy in all of our atmospheric models. Hence it is not straightforward to quantify the heat flux or $K_{\rm zz}$ directly from our results, for atmospheres with initially different $P_{\rm RCB}$. As a qualitative assessment based on the small effective spatial range of turbulence ($\sim2H$) and the rapidly decreasing heat flux outside that range, it is likely that the magnitude of the heat flux penetrating into the RCB would be insignificant, unless turbulence occurred sufficiently close to the RCB. However, considering the power-law tail of $K_{\rm zz}$ shown in Figure \ref{fig:P_Kzz} extending to higher $P$, for planets with higher entropy (lower $P_{\rm RCB}$) and with long-lived or continuously created turbulence in less dense regions, we can still consider the cumulative effects of small, but continuous heat energy supplies into deep regions.

\end{enumerate}

\section{Summary and future direction}
\label{sec:summary}

We have performed $3-$dimensional hydrodynamics simulations to investigate the effects of shock and turbulence on energy penetration into hot Jupiter atmospheres, under a variety of shear gradients. We find that the effects of turbulence on the kinetic and heat energy transfer are local, generally within a spatial range of $z\sim2H$, below the shear layer.  However, turbulence can drive a spatially and thermally great influence on in the regions above it. The temperature increases most significantly near the shear layer due to turbulence, which can further enhance the temperature inversion, in addition to the other effects already discussed in the literature \citep{Showman+2008,RauscherMenou2010}. We also find that shock formation is insignificant.
The time-averaged heat energy flux at $P\sim 1\Pbar$ when the atmosphere becomes steady is on the order of 0.001\% of $F_{\star}$ with a shear motion at the top of the atmosphere ($P_{\rm sh}\simeq 1\Pmbar$) and 0.1\% with a deeper shear layer at $P_{\rm sh}\simeq 100\Pmbar$. Accordingly, $K_{\rm zz}$ is higher for the deeper shear layer. Therefore, our results suggest that turbulence near less dense regions ($P\gtrsim 1\Pmbar$) does not lead to transport of heat energy deep enough to explain the inflated radii of hot Jupiters, regardless of how violent the turbulence is. On the other hand, as eddy motions occur at deeper regions ($P\gtrsim 100\Pmbar$), it is more likely that the heat energy is transferred more effectively throughout the atmosphere (upwards and downwards) due to relatively large kinetic energy budgets. Therefore, it is more important how deep turbulence occurs in the atmosphere (or, $P_{\rm sh}$), than how unstable the atmosphere is (or, $Ri$) for effective transfer of energy.

Understanding the role of turbulence itself is a crucial step prior to modelling global-scale atmospheres. Future work will aim at modeling global circulation of hot Jupiters including radiation.

\vspace{0.5cm}

\section*{Acknowledgements}

We are grateful to Kevin Heng and Andrew Youdin for their constructive feedbacks. We also thank the anonymous referee for constructive comments and suggestions which helped us to improve the paper. The authors acknowledge the analysis toolkit \texttt{yt} \citep{Turk+2011} and matplotlib \citep{Hunter:2007} for making the plots in the paper. The authors would like to thank Stony Brook Research Computing and
Cyberinfrastructure, and the Institute for Advanced Computational
Science at Stony Brook University for access to the high-performance
SeaWulf computing system, which was made possible by a \$1.4M National
Science Foundation grant (\#1531492).  MZ was supported by DOE/Office
of Nuclear Physics grant DE-FG02-87ER40317 for development of Castro.
Castro is freely available at
\url{https://github.com/AMReX-Astro/Castro}.  The simulations used the
    {\tt planet} setup in the code repository.











\bsp	
\label{lastpage}
\end{document}